\documentclass[journal,12pt,draftclsnofoot,onecolumn,twoside]{IEEEtran}
\usepackage{times,amsmath,epsfig,amssymb,cite,bm,float,epstopdf,graphicx,graphics,amsthm}
\usepackage{relsize}
\usepackage{dblfloatfix}
\usepackage{balance}
\usepackage[stable]{footmisc}
\usepackage{tikz,pgfplots}
\usepackage[stable]{footmisc}
\usepackage{lettrine}
\usepackage[font={footnotesize},labelfont={footnotesize}]{caption}
\usepackage{algorithm,algpseudocode}
\usepackage{tabularx,multirow,booktabs}
\usepackage{rotating,afterpage,lscape}
\usepackage{subfigure,float}
\usepackage{adjustbox}
\usepackage{enumitem}
\usepackage{afterpage}
\usepackage{chronology}

\newcommand{\C}{{\mathbb C}}
\newcommand{\R}{{\mathbb R}}
\newcommand{\N}{{\mathbb N}}
\newcommand{\E}{{\mathbb E}}
\DeclareMathOperator{\tr}{tr}

\newcommand{\bH}{{\bf H}}
\newcommand{\bB}{{\bf B}}
\newcommand{\bD}{{\bf D}}

\newcommand{\by}{{\bf y}}
\newcommand{\bx}{{\bf x}}
\newcommand{\bn}{{\bf n}}

\newcommand{\bs}{{\bf s}}
\newcommand{\bT}{{\bf T}}
\newcommand{\bb}{{\bf b}}

\tikzset{>=latex}
\newcolumntype{C}{>{\centering\arraybackslash}X} 
\title{Performance of Multibeam Very High Throughput Satellite Systems Based on  FSO Feeder Links with HPA Nonlinearity}

\author{Emna~Zedini,~\IEEEmembership{Member,~IEEE,}~Abla~Kammoun,~\IEEEmembership{Member,~IEEE,}\\~and~Mohamed-Slim~Alouini,~\IEEEmembership{Fellow,~IEEE}
\thanks{E. Zedini, A. Kammoun and M.-S. Alouini are with the Computer, Electrical, and Mathematical Science and Engineering (CEMSE) Division, King Abdullah University of Science and Technology (KAUST), Thuwal, Makkah Province, Saudi Arabia (e-mails:\{emna.zedini, abla.kammoun, slim.alouini\}@kaust.edu.sa).}
}

\begin{document}
\pgfkeys{/pgf/fpu}
\pgfmathparse{16383+1}
\edef\tmp{\pgfmathresult}
\pgfkeys{/pgf/fpu=false}

\newcommand{\boundellipse}[3]
{(#1) ellipse (#2 and #3)
}
\maketitle
\begin{abstract}
Due to recent advances in laser satellite communications technology, free-space optical (FSO) links are presented as an ideal alternative to the conventional radio frequency (RF) feeder links of the geostationary satellite for next generation very high throughput satellite (VHTS) systems. In this paper, we investigate the performance of multibeam VHTS systems that account for nonlinear high power amplifiers at the transparent fixed gain satellite transponder. Specifically, we consider the forward link of such systems, where the RF user link is assumed to follow the shadowed Rician model and the FSO feeder link is modeled by the Gamma-Gamma distribution in the presence of beam wander and pointing errors where it operates under either the intensity modulation with direct detection or the heterodyne detection. Moreover, zero-forcing precoder is employed to mitigate the effect of inter-beam interference caused by the aggressive frequency reuse in the user link. The performance of the system under study is evaluated in terms of the outage probability, the average bit-error rate (BER), and the ergodic capacity that are derived in exact closed-forms in terms of the bivariate Meijer's G function. Simple asymptotic results for the outage probability and the average BER are also obtained at high signal-to-noise ratio.
\end{abstract}


\section{Introduction}
The design of a unified platform that offers ubiquitous broadband global network coverage
with very low latency communications and very high data rates in the order of Tbit/s is increasingly becoming a challenging task for 5G and beyond 5G wireless communication systems \cite{6GNextFrontier,Debbah2019,Visionof6G,Shuping2019,survey6G}. To this end, the integration of satellite communications (SatCom), aerial networks, and terrestrial communications into a single wireless network, called space-air-ground integrated network (SAGIN), is deemed from now on crucial \cite{SAGIN}. More specifically, broadband multibeam SatCom systems are expected to provide seamless reliable and high data rate services at any place on the earth, particularly unserved and underserved areas \cite{SAGIN}. In such platform, the ground stations feed the satellite through a high capacity link, i.e. the feeder link, and then the satellite communicates the signal to different user terminals (UTs) via multiple beams, i.e. the user links \cite{LDR}.
Based on the orbit type and the altitude, satellites can be classified into three categories, namely, Low Earth Orbit (LEO), Medium Earth Orbit (MEO), and Geosynchronous Orbit (GEO) where the latter provides the greatest coverage \cite{SAGIN}.
Currently existing GEO satellite systems are based on the radio-frequency (RF) technology such as Ka-Sat with a throughput of 70 Gbit/s, Viasat 1 with 140 Gbit/s, Viasat 2 with 350 Gbit/s, and Viasat 3 that expectedly will provide throughput in the range of 1 Tbit/s by 2020 \cite{LDR}.
 In recent years, different constellations of satellites have been proposed to provide global broadband access to internet including the Starlink supported by SpaceX with 12000 LEO satellites \cite{starlink}, Oneweb with 900 LEO satellites \cite{oneweb}, and Telesat LEO with 300 to 500 satellites \cite{telesat}. All these constellations
 are based on the conventional RF solutions for both feeder and user links operating at the Ku-band (12-18 GHz), the Ka-band (27-40 Ghz), and the Q/V band (40-50 GHz). Obviously, the bandwidth limitation remains one of the key challenges when increasing the capacity. For instance, around 50 ground stations are required to reach a satellite capacity of 1 Tbit/s with the traditional RF feeder links, and the number of these ground stations increases linearly with the system throughput \cite{LDR}. Another key issue with RF links is the high risk of interference with other communication systems, leading to signal interception or jamming.
 Free-space optical (FSO) technologies are being substantially considered as an attractive alternative to the existing RF feeder links for next generation very high throughput satellite systems (VHTS) \cite{HTSEurope,Dimitrov14,Cowley14,opticalfeederlink,OpticalFeederperspectives,Roy15,Poulenard17,Toyoshima18,Toyoshima19}. By using dense-division-multiplexing (DWDM), a fiber based technique, more than one Tbit/s can be sent by a single optical ground station (OGS) to the GEO satellite, leading to the minimization of the number of required ground stations and hence drastically reducing the ground network cost \cite{OpticalFeederperspectives}. Besides the wide available bandwidth (THz) without any restriction or regulation (license-free spectrum), FSO communications hold the advantages of immunity to interference due to the very narrow laser beams along with lesser size, weight, and power compared to their RF counterparts.
 FSO systems can be classified into two categories based on the detection type at the receiver side, namely coherent and non-coherent. Non-coherent systems, also known as intensity modulation with direct-detection (IM/DD), are commonly used in FSO links mainly because of their simplicity and low cost \cite{Sandalidis}. In such systems, the receiver directly detects the intensity of the emitted light. With recent advances in integrated circuits as well as high-speed digital signal processing, coherent detection is becoming more attractive \cite{coherentmod4,coherentmod5,SpectrtalEfficiency}.
In such systems, the incoming optical signal is mixed with a local oscillator (LO) before photo-detection, which improves the receiver sensitivity \cite{ReceiverSensistivity}. Another interesting property of coherent detection is that amplitude, frequency, and phase modulation can be employed, which considerably increase the system spectral efficiency \cite{SpectrtalEfficiency}. Furthermore, coherent detection allows background noise rejection \cite{coherentmod3}.
Although most of laser satellite communication
(laser SatCom) systems, currently under development,
are operating using the IM/DD technique, coherent detection
systems have also been employed as a viable alternative for
certain applications \cite{Andrews,Toyoshima11,Surof17}.

On the other hand, the primary concerns of the FSO feeder link are atmospheric turbulence, beam wander, and misalignment pointing errors.
The atmospheric turbulence is caused by fluctuations in the refractive index resulting in strong intensity fluctuations, or scintillations, that may cause severe performance degradation of the FSO link. The scintillation index, i.e. normalized variance of the irradiance, is generally used to characterize these irradiance fluctuations.
As for the beam wander, it is caused by deviations of the beam from the boresight due to the presence of turbulent eddies larger than the beam diameter. This beam wander effect can hence lead to strong fading of the received signal \cite{Andrews}.

Lastly, maintaining a constant line-of-sight (LOS) communication between the transceivers is very essential to have a 100\% availability of the FSO feeder link, where the optical beam is highly directional with very narrow beam divergence. Due to the satellite mechanical vibration \cite{Chen89,Arnon97,Li12,Dong2015,KaushalSurvey}, the transmitted beam to the receiver satellite vibrates leading to a misalignment between the transmitter and the receiver, known also as pointing error. These pointing errors may lead to significant performance degradation or result in failure of the FSO link \cite{KaushalSurvey,Najafi18}.

Despite all these technical challenges, the FSO feeder-link remains the most promising technological option for next generation VHTS systems\cite{opticalfeederlink,LDR}. For instance, in the frame of the Terabit-throughput optical satellite system technology (THRUST) project, the German Aerospace Agency DLR set the world-record in FSO communications to 1.72 Tbit/s and 13.16 Tbit/s in 2016 and 2017, respectively \cite{THRUST,OpticalFeederExperiment}.

On the RF user link side, full frequency reuse is employed where all beams operate at the same frequency in order to enhance the bandwidth efficiency of the system. However, such an aggressive use of the spectrum introduces inter-beam interference, i.e. each UT receives interference from adjacent beams. An efficient interference mitigation technique consists of precoding the signals at the OGS before transmitting them to the different UTs, but requires the availability of accurate channel state information (CSI) at the OGS \cite{robust,multicast,JoroughiMultibeam,precoding16}. However, acquiring up to date and reliable CSI at the OGS introduces a long delay, leading to an outdated CSI. Subsequently, the slow fading channel in fixed satellite services (FSS), where the UTs have fixed positions inside the beams \cite{multicast,Taricco,ErichBook}, would facilitate
the CSI acquisition process as CSI needs to be updated less
frequently.
Interestingly, in \cite{CapacityZF}, a novel zero-forcing (ZF) precoding scheme has been proposed for FSS multibeam SatCom systems that only exploits the UTs positions and the antenna beam radiation pattern, without requiring any CSI at the OGS. More precisely, the OGS can generate the deterministic multibeam matrix without requiring any feedback from the UTs \cite{Christopoulos,Zheng2012,multicast,LetzepisMutibeam,SatelliteFSO,CapacityZF,Arnau2014}.

The vast majority of conventional high power satellite transponders employ travelling
wave tube amplifier (TWTA) and solid-state power amplifier (SSPA) as onboard memoryless high power amplifiers (HPAs).
For high output powers, TWTAs are commonly
employed, in particular at higher frequency bands, because they offer higher data rates and greater bandwidth with better efficiency than SSPAs.
For lower frequency bands and for
lower transmitter power applications, SSPAs are generally preferred as they exhibit higher reliability, lower mass, and
better linearity \cite{ErichBook,TWTASSPA}.
However, these HAPs models have two major nonlinear characteristics, namely, amplitude to amplitude modulation (AM/AM) and amplitude to phase modulation (AM/PM) conversions that should be taken into account as they can lead to severe performance degradation. Therefore, several models have been proposed to represent these nonlinear characteristics, mainly, Saleh model \cite{Saleh} and Rapp model \cite{Rapp} to characterize the nonlinear distortion due to TWTA and SSPA models, respectively. A few works have studied the impact of hardware impairments on the performance of satellite relay networks \cite{satelliteNL1,satelliteNL2,satelliteNL3}. Their results demonstrate that the impairments degrade the system performance, in particular when the impairments level is larger.

Motivated by the DLR experimental demonstration in \cite{OpticalFeederExperiment}, we propose in this work to
investigate the performance of VHTS FSO systems with multi-beam RF capabilities.
As per authors' best knowledge, the first
performance analysis of multibeam high throughput satellite systems with optical
feeder links has been carried out in \cite{SatelliteFSO}. More specifically, the FSO feeder link is considered to be operating using direct detection over the lognormal distribution whose scope is restricted to weak turbulence channel conditions, and the RF user link is modeled by the double-lognormal fading. Based on the precoding scheme presented in \cite{CapacityZF}, the authors provided approximations for the outage probability,
the average BER for MQAM and MPSK modulation schemes, and the ergodic capacity in the case of linear power amplifier (PA) at the fixed gain satellite transponder. However, to the best of our knowledge, there are no exact closed form expressions that capture the outage probability, the average BER for a variety of modulation schemes, and the ergodic capacity performance under both IM/DD and heterodyne detection techniques with HPA nonlinearity taken into account.
In this context, this work presents, for the first time, a unified analytical framework for the calculation of the fundamental performance metrics of multibeam VHTS systems with HPA nonlinearity in exact closed form, applicable to both types of detection techniques.
The FSO feeder link is modeled by the Gamma-Gamma distribution, a good model
for atmospheric turbulence under both small and large scales
atmospheric fluctuations \cite{andrews1}, in the presence of beam wander and pointing errors. On the other hand, the RF user links are modeled as shadowed Rician channels that have been proposed in \cite{LMSAlouini} for land mobile satellite channels (LMS). Indeed, it has been shown in \cite{LMSAlouini} that the shadowed Rician model provides an excellent fit to the experimental data and has a simple mathematical form, making it attractive from a performance analysis point of view. Hence, the main contributions of this work are stated as follows.
\begin{itemize}
\item We present a detailed description of the system and channel models with a particular focus on the statistics of the FSO feeder link to stress that there is a great difference between modeling horizontal propagation paths and slant paths, where it is required to consider changes in the refractive index structure parameter along the path.

\item We introduce TWTA and SSPA nonlinear amplifiers along with their impairment parameters, and utilize the Bussgang linearization theory to linearize the distortion introduced by these two HPAs.

\item We first derive the end-to-end signal-to-noise-plus-distortion-ratio (SNDR) in the case of fixed gain transparent satellite transponders, considering both types of detection techniques for the FSO feeder link (i.e. IM/DD and heterodyne) and using the ZF precoder proposed in \cite{CapacityZF}.

\item  Capitalizing on this result, we present closed-form expressions for the cumulative distribution function (CDF) and the probability density function (PDF) in terms of the bivariate Meijer's G function, and the moments in terms of simple functions.

\item We then derive the outage probability, the average bit-error rate (BER) of a variety of modulation schemes,
and the ergodic capacity, all in terms of the bivariate Meijer's G function.

\item Finally, we present very tight asymptotic expressions for the outage probability and the average BER in the high signal-to-noise ratio (SNR) region in terms of simple elementary functions which are particularly useful to reveal some physical insights.
\end{itemize}

The remainder of this paper is organized as follows. The system and channel models are outlined in Section II. We derive the statistics of the end-to-end SNDR in Section III and we present closed-form expressions for the performance metrics along with the asymptotic results at high SNR regime in Section IV.
Numerical and simulation results are then provided in Section V followed by the conclusions in
Section VI.
\section{System and Channel Models}
\begin{figure}[!h]
    \centering
\includegraphics[scale=1]{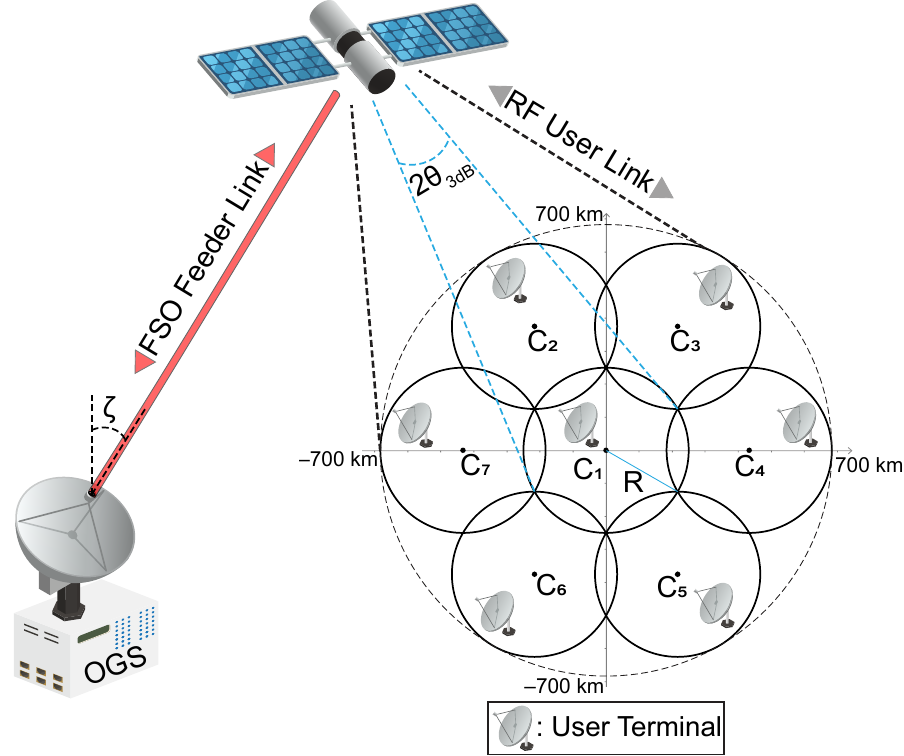}
\caption{Architecture of a multibeam very high throughput GEO satellite system with FSO feeder links.}
\label{fig:systemmodel}
\end{figure}

We consider the forward link of a multibeam VHTS system which is defined as the end-to-end link from the OGS to the different UTs.
More specifically, it includes the uplink of the feeder link (i.e. the link between the OGS and the GEO satellite), the transparent or non-regenerative GEO satellite with $N$ antenna feeds, and the downlink of the user link (i.e. the link between the GEO satellite and the UTs). In addition, we consider that the feeder link is a high capacity FSO single-input
single-output (SISO) link, whereas the user link is a multiuser multiple-input single-output (MISO) Ka-band RF link as shown in Fig.~\ref{fig:systemmodel}. In this paper, we assume a high-energy FSO link whose performance is limited by shot noise as well as thermal noise. In this case, the noise can be modeled to high accuracy as zero mean, signal independent additive white Gaussian noise (AWGN) (a widely accepted assumption in many reported works in the literature \cite{AWGN1,shotnoise1,shotnoise2,murat}.
\noindent
In this context, a single OGS simultaneously serves multiple UTs via $N$ beams, in a single feed per beam scenario. Moreover, on the user link side, full-frequency reuse is assumed with a cluster size $K=1$, where all beams operate at the same frequency. The coverage area of the GEO satellite is filled up with seven beams arranged in a circular way, resulting in overlapping regions as detailed in Fig.~\ref{fig:systemmodel}. With beam radius $R$, the coordinates of each beam center are determined by $C_1(0,0)$, $C_2\left ( -\frac{\sqrt{3} }{2}R,\frac{3}{2}R\right )$, $C_3\left ( \frac{\sqrt{3} }{2}R,\frac{3}{2}R\right )$, $C_4\left ( \sqrt{3}R,0\right )$, $C_5\left ( \frac{\sqrt{3}}{2}R,-\frac{3}{2}R\right )$, $C_6\left ( -\frac{\sqrt{3}}{2}R,-\frac{3}{2}R\right )$, and $C_7\left ( -\sqrt{3}R,0\right )$.
Furthermore, we focus herein on FSS systems and therefore the UTs have fixed positions inside the beams and can be ultra small aperture terminals (USATs) \cite{ErichBook}, usually equipped with single antennas, as shown in Fig.~\ref{fig:systemmodel}.
\subsection{FSO Feeder Link}
We assume that the data is first precoded at the OGS before being transmitted in order to mitigate inter-beam interference.
Moreover, DWDM techniques are used to provide the aggregated throughput of
multiple Tbit/s where the optical carriers, modulated using either intensity modulation or coherent modulation, are multiplexed into a single-mode fiber (SMF), amplified, and sent through the telescope of the OGS towards the GEO satellite.
At the GEO satellite, the optical signal is captured by the telescope, demultiplexed to separate the individual DWDM channels, converted to electrical RF channels in the Ka-band, and sent to the users \cite{OpticalFeederperspectives}.
Then, the received signal at
the GEO satellite, $\by_1 \in \C^{N\times 1}$, can be expressed as
\begin{align}\label{y1}
\by_1= \eta I \bx +\bn_1,
\end{align}
where $\eta$ stands for the effective photoelectric
conversion ratio, $I$ represents the received optical irradiance, $\bx \in \C^{N\times 1}$  is the precoded transmit signal vector with a total power constraint of $\E[\bx \bx^{\mbox{\tiny H}} ] \leq  P_g$, and $\bn_1 \in \C^{N\times 1}$ refers to the additive noise vector consisting of circularly symmetric complex Gaussian entries with zero-mean and variance $\sigma_1^2$, i.e. $\mathcal{CN}(0,\sigma_1^2)$. The irradiance $I$ includes the effect of the path loss $I_l$, the attenuation caused by the atmospheric turbulence $I_a$, and the attenuation due to pointing errors $I_p$, i.e. $I=I_l I_a I_p$. The path loss $I_l$ is deterministic and is described by the exponential Beers-Lambert Law as $I_l=\exp(- \sigma L)$ where $\sigma$ represents the atmospheric attenuation coefficient and $L$ is the FSO link length \cite{Steve07}.

The pointing error loss due to misalignment is caused by the displacement of
the laser beam along elevation and azimuth
directions that are typically modeled as independent and identically
distributed Gaussian random variables with zero mean value and variance $\sigma_s^2$.
The resulting radial displacement at the receiver $r$ is therefore statistically characterized by a Rayleigh distribution \cite{Chen89,Li12,KaushalSurvey,Steve07} for which the PDF of the irradiance $I_p$ is given by \cite[Eq.(11)]{Steve07}
 \begin{align}\label{PDFofPE}
 f_{I_p}(I_p)=\frac{\xi ^{2}}{A_0^{\xi ^{2}}}\,I_p^{\xi ^{2}-1},\quad 0\leq I_p \leq A_0,
 \end{align}
 where $A_0$ is the fraction of the collected power at $r=0$ and $\xi$ is defined as the ratio between the equivalent beam radius at the receiver and the jitter standard deviation at the receiver, and used to quantify the severity of the pointing error effect \cite{Steve07}.

 The atmospheric turbulence $I_a$ is modeled by the Gamma-Gamma distribution whose PDF is given in \cite{Andrews} as
\begin{align}\label{PDFofIa}
f_{I_a}(I_a)=\frac{2 (\alpha \beta)^{\frac{\alpha+\beta}{2}}}{\Gamma(\alpha)\Gamma(\beta)} I_a^{\frac{\alpha+\beta}{2}-1}K_{\alpha-\beta}\left ( 2 \sqrt{\alpha \beta I_a} \right ), \quad I_a> 0
\end{align}
where $\Gamma(\cdot)$ represents the gamma function \cite[Eq.(8.310/1)]{Tableofintegrals}, $K_{\alpha-\beta}(\cdot)$ stands for the
modified Bessel function of the second kind with order $\alpha-\beta$, and $\alpha$ and $\beta$ are positive parameters which are related to the large- and
small-scale irradiance fluctuations, respectively. Taking into account the effect of beam wander, the parameters $\alpha$ and $\beta$ for an untracked collimated beam are defined as \cite[p.~517]{Andrews}
\begin{align}\label{alphaEq}
\nonumber \alpha &=\left[5.95 (H-h_0)^2 \sec^2(\zeta)\left (\frac{ 2 W_0}{r_0} \right )^{\frac{5}{3}} \left ( \frac{\alpha_{\rm{pe}}}{W} \right )^2\right.\\
&\left.+\exp\left( \frac{0.49 \sigma_{\rm{Bu}}^2 }{\left ( 1+0.56 \sigma_{\rm{Bu}}^{\frac{12}{5}} \right )^{\frac{7}{6}}}\right)-1  \right ]^{-1}
\end{align}
\noindent and
\begin{align}\label{betaEq}
\beta=\left[\exp\left( \frac{0.51 \sigma_{\rm{Bu}}^2 }{\left ( 1+0.69 \sigma_{\rm{Bu}}^{\frac{12}{5}} \right )^{\frac{5}{6}}}\right)-1  \right ]^{-1},
\end{align}
where $H$ represents the altitude of the GEO satellite in m, $h_0$ is the altitude of the optical ground station in m, $\zeta$ refers to the zenith angle, $W_0$ denotes the beam radius at the transmitter, $W$ is the beam radius at the receiver $W=W_0\sqrt{\Theta_0^2+\Lambda_0^2}$ where $\Theta_0$ and $\Lambda_0$ are the transmitter beam parameters defined as $\Theta_0^2=1-L/F_0$ and $\Lambda_0=2 L/(k\, W_0^2)$ with $F_0$ being the phase front radius of curvature at the transmitter ($F \to \infty$ for a collimated beam), $L=(H-h_0) \sec(\zeta)$, $k= 2\pi/\lambda$ denotes the wavelength number, $r_0$ stands for the Fried parameter, $\alpha_{\text{pe}}$ is the beam wander-induced angular pointing error and $\sigma_{\rm{Bu}}^2$ is the Rytov variance given by \cite[Eq.(8)]{Subrat2015}

\begin{align}\label{Rytovvar}
\nonumber \sigma_{\rm{Bu}}^2&=2.25\,k^{\frac{7}{6}}(H-h_0)^{\frac{5}{6}}\sec^{\frac{11}{6}}(\zeta)\int_{h_0}^{H}C_n^2(h)\\
& \times \left ( 1-\frac{h-h_0}{H-h_0} \right )^{\frac{5}{6}}\left (\frac{h-h_0}{H-h_0} \right )^{\frac{5}{6}}dh.
\end{align}\noindent
The Fried parameter $r_0$ is defined as \cite[p.~492]{Andrews}
\begin{align}\label{Fried}
r_0=\left [0.42\, \sec(\zeta)\,k^2 \int_{h_0}^{H}C_n^2(h)\,dh  \right ]^{-\frac{3}{5}},
\end{align}
where $C_n^2(h)$ is the refractive index structure parameter that varies as a function of the altitude $h$ based on the most widely used Hufnagel-Valley
(H-V) model as \cite[p.~481]{Andrews}
\begin{align}\label{Cn2}
\nonumber C_n^2(h)&=0.00594 \left ( \frac{w}{27} \right )^2 \left ( 10^{-5}h \right )^{10}\exp\left ( -\frac{h}{1000} \right )\\
&+2.7 \times 10^{-16}\exp\left ( -\frac{h}{1500} \right )+C_n^2(0) \exp\left ( -\frac{h}{100} \right ),
\end{align}
where $w$ denotes the rms windspeed in m/s and $C_n^2(0)$ refers to the ground level turbulence in m$^{-2/3}$.

The beam wander-induced pointing error variance $\sigma_{\rm{pe}}^2$ is related to $\alpha_{\rm{pe}}$ such that $\alpha_{\rm{pe}}=\sigma_{\rm{pe}}/L$ and can be expressed as \cite[p.~503]{Andrews}
\begin{align}\label{PEvar}
\sigma_{\rm{pe}}^2&=0.54\,(H-h_0)^2 \sec^2(\zeta) \left (\frac{ \lambda }{2 W_0}\right )^2 \left (\frac{2 W_0}{r_0}  \right )^{\frac{5}{3}}
\left [ 1-\left (\frac{C_r^2 W_0^2/r_0^2}{1+C_r^2 W_0^2/r_0^2}  \right ) ^{\frac{1}{6}}\right ],
\end{align}
where $C_r$ is a scaling constant set as $2 \pi$ \cite{Subrat2015} and $r_0$ given by (\ref{Fried}).

Based on (\ref{PDFofPE}), (\ref{PDFofIa}), and the path loss expression, the PDF of $I=I_l I_a I_p$ under the combined effect of atmospheric turbulence, beam wander, pointing errors, and path loss can be written as

\begin{align}\label{PDFofI}
f_{I}(I)=\frac{\xi^2 \alpha \beta}{A_0 I_l\Gamma(\alpha)\Gamma(\beta)}\, {\rm{G}}_{1,3}^{3,0}\left[ \frac{\alpha \beta }{A_0 I_l}I \left| \begin{matrix} {\xi^2} \\ {\xi^2-1,\alpha-1,\beta-1} \\ \end{matrix} \right. \right],
\end{align}
where ${\rm{G}}_{p,q}^{m,n}(\cdot)$ is the Meijer's G function \cite[Eq.(9.301)]{Tableofintegrals}.

\subsection{Nonlinear Satellite Transponder}
Considering nonlinear HPA at the satellite transponder, the amplification process is performed in two distinct steps. In the first step, the RF precoded
signal $\by_1$ is amplified with a constant gain matrix ${\bf F}=G{\bf I}_N$, that is $\by_s={\bf F} \by_1$, where the amplification factor $G=\sqrt{\frac{P_r}{P_g \,\mathbb{E} \left [(\eta I)^r  \right ]+\sigma_1^2}}$ is selected such that the total transmit power constraint at the satellite transponder is met, i.e. $\mathbb{E}\left [ \left \| {\bf F} \by_1\right \| ^2 \right ]\leq P_r$, where $P_r$ is the mean signal power at the output of the gain block.

In the second phase, the amplified version of the signal $\by_s$ is passed through a nonlinear circuit and then the signal at the output of the memoryless nonlinear HPA can be given as \cite{Aissa10}
\begin{align}\label{ysnl1}
\by_{s_{\rm{NL}}}=f_A(\left \|\by_s \right \|)\exp\left( j\left(f_P(\left \|\by_s \right \|)+{\rm{arg}}(\by_s)\right)\right),
\end{align}
where $f_A(\cdot)$ and $f_P(\cdot)$ represent the AM/AM and AM/PM characteristic functions, respectively, and $j^2=-1$. As discussed in Section I, we consider two types of nonlinear amplifiers which are widely employed in conventional high power satellite transponders, namely TWTA and SSPA.
For the TWTA model, the AM/AM and AM/PM conversions are given as \cite{Saleh}
\begin{align}
f_A(\left \| \by_s \right \|)=A_{\rm{sat}}^2 \frac{\left \| \by_s \right \|}{\left \| \by_s \right \|^2+A_{\rm{sat}}^2};\quad f_P(\left \| \by_s \right \|)={\rm{\Phi_0}}  \frac{\left \| \by_s \right \|}{\left \| \by_s \right \|^2+A_{\rm{sat}}^2},
\end{align}
where $A_{\rm{sat}}$ represents the input saturation amplitude level and ${\rm{\Phi_0}}$
controls the maximum phase distortion introduced by TWTA. The AM/AM and AM/PM functions of the SSPA model
can be expressed as \cite{Rapp}
\begin{align}
f_A(\left \| \by_s \right \|)= \frac{\left \| \by_s \right \|}{\left [\left (\frac{\left \| \by_s \right \|}{A_{\rm{sat}}}  \right )^{2v}+1  \right ]^{\frac{1}{2v}}};\quad f_P(\left \| \by_s \right \|)=0,
\end{align}
where $v$ refers to the smoothness factor that controls the transition
from linear to saturation region.

It is worthy to mention that the nonlinear distortion created by both TWTA and SSPA models as shown by (\ref{ysnl1}) makes it very hard to obtain closed-form and easy-to-use expressions for
important performance metrics such as the outage probability
and the average BER of the VHTS FSO system under consideration. However, we can linearize this distortion by means of using the Bussgang's
linearization theory \cite{Bussgang} since the input signal $\by_s$ can be approximately modeled as Gaussian distributed. This is due to the fact that the precoded transmit signal $\bx$ given in (\ref{precodedx}) is a weighted sum of independent and identically distributed (i.i.d.) random variables that can be approximated by a Gaussian distribution according the central limit theorem.
By using the Bussgang's theorem, the output of the nonlinear HPA can be expressed as
\begin{align}\label{ysnl}
\by_{s_{\rm{NL}}}=K \by_s +\bn_{\rm{NL}},
\end{align}
where $K$ is the linear scale parameter and $\bn_{\rm{NL}} \in \C^{N\times 1}$ is the nonlinear distortion term uncorrelated with $\by_s$ and modeled as $\mathcal{CN}(0,\sigma_{\rm{NL}}^2)$.
As discussed in Section I, we consider two types of nonlinear amplifiers which are widely employed in conventional high power satellite transponders, namely TWTA and SSPA. More specifically, in the case of TWTA model, the impairment parameters $K$ and $\sigma_{\rm{NL}}^2$ can be given under the assumption of negligible AM/PM effects by \cite[Eq.(11)]{NLHPA}, \cite[Eq.(19)]{Balti2017}
\begin{equation}\label{TWTAparam}
\begin{aligned}
&K=\sqrt{\frac{A_{\rm{sat}}^2}{4P_r}}\left [ \sqrt{\frac{4A_{\rm{sat}}^2}{P_r}} -\sqrt{\pi }\exp\left ( \frac{A_{\rm{sat}}^2}{P_r} \right ){\rm{erfc}}\left ( \sqrt{\frac{A_{\rm{sat}}^2}{P_r}} \right )\left ( \frac{2A_{\rm{sat}}^2}{P_r} -1\right )\right ]\\
&\sigma_{\rm{NL}}^2=A_{\rm{sat}}^2 \left [  1+\frac{A_{\rm{sat}}^2}{P_r} \exp\left ( \frac{A_{\rm{sat}}^2}{P_r} \right ){\rm{Ei}}\left ( -\frac{A_{\rm{sat}}^2}{P_r} \right )\right ]-K^2 P_r,
\end{aligned}
\end{equation}
where $A_{\rm{sat}}$ represents the input saturation amplitude level, ${\rm{erfc}}(\cdot)$ is the complementary error function \cite[Eq.(8.250/4)]{Tableofintegrals}, and ${\rm{Ei}}(\cdot)$ is the exponential integral function \cite[Eq.(8.21)]{Tableofintegrals}.
Additionally, for the SSPA model, $K$ and $\sigma_{\rm{NL}}^2$ can be expressed using \cite[Eq.(18)]{Balti2017} as
\begin{equation}\label{SSPAparam}
\begin{aligned}
&K=\frac{A_{\rm{sat}}^2}{P_r}\left [ 1+\frac{A_{\rm{sat}}^2}{P_r} \exp\left ( \frac{A_{\rm{sat}}^2}{P_r} \right ){\rm{Ei}}\left ( -\frac{A_{\rm{sat}}^2}{P_r} \right )\right ]\\
&\sigma_{\rm{NL}}^2=-\frac{A_{\rm{sat}}^4}{P_r} \left [ \left ( 1+\frac{A_{\rm{sat}}^2}{P_r} \right ) \exp\left ( \frac{A_{\rm{sat}}^2}{P_r} \right ){\rm{Ei}}\left ( -\frac{A_{\rm{sat}}^2}{P_r} \right )+1\right ]-K^2 P_r.
\end{aligned}
\end{equation}
It is important to mention that in practice, the satellite transponders are not operated at saturation but backed off in order to reduce the nonlinearity. Indeed, the actual transmission power is reduced by a given amount below the HPA saturation point, which is known as the input back-off (IBO) and is defined as \cite{NLHPA}
\begin{align}
{\rm{IBO}}=\frac{A_{\rm{sat}}^2}{P_r}.
\end{align}
Moreover, it is noteworthy that $K$ and $\sigma_{\rm{NL}}^2$ depend on IBO for both TWTA and SSPA models, and are constants for a fixed IBO value.

\subsection{RF User Link}
The received signal vector at all the UTs can be given as
\begin{align}\label{y2}
\by_2=\bH \by_{s_{\rm{NL}}} +\bn_2=K G \eta I \bH \,\bx+K G \bH \bn_1+\bH \bn_{\rm{NL}}+\bn_2,
\end{align}
where the user link channel matrix $\bH \in \C^{N\times N}$ represents the channel gains between the $N$ feeds and the $N$ UTs and takes into account the atmospheric fading, the beam radiation pattern, and path losses, and $\bn_2 \in \C^{N\times 1}$ refers to the noise vector with elements drawn from $\mathcal{CN}(0,1)$. Using \cite{JoroughiMultibeam,SatelliteFSO,CapacityZF}, the user link channel matrix can be expressed as
\begin{align}\label{matrixH}
\bH=\bD \bB,
\end{align}
where $\bB \in \R^{N\times N}$ is the multibeam gain matrix that models the satellite antenna radiation pattern, the receive antenna gain, and
the path loss \cite{Arnau2014,SatelliteFSO}.
Assuming that the receive antenna gains for all UTs are identical and equal to $G_t$, the transmitter antenna gains for all satellite feeds are identical and equal to $G_r$, neglecting the Earth curvature so that all UTs have a common slant range equal to the GEO satellite elevation distance,
and employing the Bessel function model for a typical tapered-aperture antenna, the beam gain
from the $j$-th feed towards the $i$-th UT can be expressed as \cite{Arnau2014,SatelliteFSO}
\begin{align}\label{Bij}
[\bB]_{ij}= \frac{c\sqrt{G_t G_r }}{4 \pi  f D \sqrt{\kappa_B T_r B_w}} \left (\frac{{\rm J}_1 (u_{ij})}{2 u_{ij}}+36 \frac{{\rm J}_3 (u_{ij})}{u_{ij}^3}\right ),
\end{align}
where $c$ is the speed of light, $f$ is the carrier frequency, $\kappa_B$ refers to the Boltzman constant, $T_r$ denotes the receiver noise temperature, $B_w$ stands for the bandwidth of the user link, ${\rm J}_1(\cdot)$ and  ${\rm J}_3(\cdot)$ are the Bessel functions of the first kind of order 1 and 3, respectively. In (\ref{Bij}), $u_{ij}=2.07123\sin(\theta_{ij})/\sin(\theta_{3\rm{dB}})$ is a function of the off-axis angle with respect to the beam's boresight $\theta_{ij}=\arctan(d_{ij}/D)$ where $d_{ij}$ represents the distance between the $i$-th UT and the $j$-th beam boresight (slant-range), $D$ is the distance from the UT to the satellite, and $\theta_{3\rm{dB}}=R/D$ refers to the beam's 3 dB angle, with $R$ denoting the beam radius.
Since we consider that UTs have fixed positions on earth, the beam gain between the $i$-th UT and the $j$-th satellite feed is fixed and therefore the multibeam gain matrix $\bB$ is deterministic \cite{LetzepisMutibeam,SatelliteFSO,CapacityZF}.

In (\ref{matrixH}), $\bD \in \C^{N\times N} $is a diagonal matrix that represents the fading in the user link with the diagonal entry $d_i$ referring to
the fading gain for the $i$-th UT, $i \in \N$ which is assumed to follow the shadowed Rician
model for LMS channels with the PDF given in \cite{LMSAlouini} by
\begin{align}\label{LMSfading}
 f_{\left |d_i  \right |}(y)&=\left (\frac{2b_i m_i}{2 b_i m_i+\Omega_i}  \right )^{m_i}\frac{y}{b_i}\exp\left ( -\frac{y^{2}}{2b_i} \right )
 _{1}F_{1}\left ( m_i,1,\frac{\Omega_i y^{2}}{2 b_i\left ( 2 b_i m_i+\Omega_i \right )} \right ), y\geq 0,
\end{align}
where $\Omega_i$ refers to the average power of the line-of-sight (LOS) component, $2b_i$ represents the average power of the multipath component, $m_i$ stands for the fading severity parameter, and $_{1}F_{1}(\cdot,\cdot,\cdot)$ is the confluent hypergeometric function \cite[Eq.(9.210/1)]{Tableofintegrals}. For $m_i=0$, (\ref{LMSfading}) reduces to the Rayleigh PDF, while for $m_i \to \infty$ it simplifies to the Rice PDF.

\subsection{Zero-Forcing Precoding}
The transmit precoded signal can be expressed as
\begin{align}\label{precodedx}
\bx=\bT \bs,
\end{align}
where $\bT \in \C^{N\times N}$ is the precoding matrix and $\bs \in \C^{N\times 1}$ represents the UTs data symbols at the OGS with $\E[\bs \bs^{\mbox{\tiny H}} ]={\bf I}_N$ and the transmit power constraint can be re-written as
\begin{align}
\E\left [ \left \| \bx \right \| ^2\right ]=\E\left [ \left \| \bT \bs \right \| ^2\right ]=\tr\left(\bT \bT^{\mbox{\tiny H}}  \right)\leq  P_g.
\end{align}
Substituting (\ref{matrixH}) and (\ref{precodedx}) into (\ref{y2}), the received signal vector can be expressed as

\begin{align}\label{y2precoding}
\by_2=K G \eta I \bD \bB \bT \bs+K G \bD \bB \bn_1+\bD \bB \bn_{\rm{NL}}+\bn_2.
\end{align}
Using the ZF precoding technique presented in \cite{CapacityZF} which does not require CSI at
the OGS and is only based on the deterministic multibeam matrix $\bB$, the precoding matrix $\bT$ can be given as
\begin{align}\label{ZFmatrix}
\bT=\sqrt{c_{\rm{ZF}} }\bB^{\mbox{\tiny H}} \left(\bB \bB^{\mbox{\tiny H}} \right)^{-1},
\end{align}
where $c_{\rm{ZF}}$ is set such that \cite{CapacityZF,SatelliteFSO}
\begin{align}\label{cZF}
c_{\rm{ZF}}=\frac{P_g}{\tr\left [ \left(\bB \bB^{\mbox{\tiny H}} \right)^{-1} \right ]}.
\end{align}
By plugging (\ref{ZFmatrix}) into (\ref{y2precoding}), the received signal at the $i$-th UT simplifies to
\begin{align}\label{y2user}
\by_{2,i}=\sqrt{c_{\rm{ZF}}} K G \eta I d_i s_i+K G d_i \bb_i^{\mbox{\tiny T}} \bn_1+d_i \bb_i^{\mbox{\tiny T}} \bn_{\rm{NL}}+\bn_{2,i}.
\end{align}
Finally, the end-to-end SNDR at the $i$-th UT can be expressed after some manipulations as
\begin{align}\label{SNDR}
\gamma_i=\frac{1}{\left \| \bb_i^{\mbox{\tiny T}} \right \|^2} \frac{\gamma_1 \gamma_{2,i}}{\kappa \gamma_{2,i}+\tr\left [ \left(\bB \bB^{\mbox{\tiny H}} \right)^{-1} \right ]\overline{\gamma}_1+\kappa},
\end{align}
where $\gamma_1=\frac{P_g (\eta I)^r}{\sigma_1^2 \tr\left [ \left(\bB \bB^{\mbox{\tiny H}} \right)^{-1} \right ]}$ is the electrical SNR of the FSO feeder link operating under either IM/DD (i.e. $r=2$) or heterodyne detection (i.e. $r=1$), $\gamma_{2,i}=\frac{P_s \left | d_i \right |^2 \left \| \bb_i^{\mbox{\tiny T}} \right \|^2}{N }$ is the SNR of the $i$-th UT, $\frac{P_s}{N}$ is the average transmitted power at the satellite satisfying $\frac{P_s}{N}=K^2 P_r+\sigma_{\rm{NL}}^2$, and $\kappa$ is the ratio between the average received SNR and
the average transmitted SNDR at the relay given by \cite{NLHPA}
\begin{align}\label{kappa}
\kappa=1+\frac{\sigma_{\rm{NL}}^2}{K^2 G^2 \sigma_1^2}.
\end{align}
Note that the parameter $\kappa$ in (\ref{kappa}) plays a key role in this paper as it describes the level of impairments, under both TWTA and SSPA models.
In addition, when $\kappa=1$, (\ref{SNDR}) reduces to the end-to-end SNR in the case of linear PA at the satellite transponder as it implies that $\sigma_{\rm{NL}}^2=0$.

Considering both IM/DD and heterodyne detections, the PDF of $\gamma_1$ can be obtained from (\ref{PDFofI}) as
\begin{align}\label{SNR1}
f_{\gamma_1}(\gamma_1)=\frac{\xi^2}{r\Gamma(\alpha)\Gamma(\beta)\gamma_1}\, {\rm{G}}_{1,3}^{3,0}\left[ \frac{\alpha \beta\, \xi^2}{(\xi^2+1)}\left ( \frac{\gamma_1}{\mu_r} \right )^{\frac{1}{r}} \left| \begin{matrix} {\xi^2+1} \\ {\xi^2,\alpha,\beta} \\ \end{matrix} \right. \right],
\end{align}
where $\mu_r$ is the average electrical SNR given by $\mu_r=\frac{P_g}{\sigma_1^2 \tr \left [  \left(\bB \bB^{\mbox{\tiny H}} \right)^{-1} \right ]}
\left (\eta A_0 I_l\xi^2/(\xi^2+1)  \right )^r$ and can be written in terms of the average SNR of the FSO feeder link, $\overline{\gamma}_1$, as
\begin{align}
\mu_r=\frac{(\xi^2+r)\left (\alpha \beta \xi^2  \right )^r \Gamma(\alpha)\Gamma(\beta)}{\xi^2 (\xi^2+1)^r\,\Gamma(\alpha+r)\Gamma(\beta)}\overline{\gamma}_1.
\end{align}\noindent
Moreover, using (\ref{LMSfading}), the PDF of the SNR $\gamma_{2,i}$ can be given as
\begin{align}\label{SNR2}
 f_{\gamma_{2,i}}(\gamma_2)&=\frac{m_i}{\overline{\gamma}_{i,2}}   \left (\frac{2b_i m_i}{2 b_i m_i+\Omega_i}  \right )^{m_i-1}\exp\left ( -\frac{(2 b_i m_i+\Omega_i)\gamma_2}{2b_i\, \overline{\gamma}_{2,i}} \right )
 _{1}F_{1}\left ( m_i,1,\frac{\Omega_i \gamma_2}{2 b_i \overline{\gamma}_{2,i}} \right ),
\end{align}
where $\overline{\gamma}_{2,i}=\frac{P_s  \left \| \bb_i^{\mbox{\tiny T}} \right \|^2}{N } (2 b_i m_i+\Omega_i)$ is the average SNR at the $i$-th RF user link.
For integer values of the fading parameter, i.e. $m_i \in \mathbb{N}$, the PDF expression in (\ref{SNR2}) can be simplified by utilizing \cite[Eq.(07.20.03.0009.01)]{Wolfram} then \cite[Eq.(07.02.03.0014.01)]{Wolfram} as
\begin{align}\label{SNR2integer}
\nonumber f_{\gamma_{2,i}}(\gamma_2)&=\frac{m_i}{\overline{\gamma}_{i,2}}   \left (\frac{2b_i m_i}{2 b_i m_i+\Omega_i}  \right )^{m_i-1}\exp\left ( -\frac{m_i\gamma_2}{\overline{\gamma}_{2,i}} \right )\\
&\times \sum_{k=0}^{m_i-1}\frac{(-1)^k (1-m_i)_k}{k!^2} \left (\frac{\Omega_i \gamma_2}{2 b_i \overline{\gamma}_{i,2}}  \right )^k,
\end{align}
where $(a)_k=\Gamma(a+k)/\Gamma(a)$ denotes the Pochhammer symbol \cite[p.~xliii]{Tableofintegrals}.
\section{Statistical Analysis}
In this section, we derive new exact closed-form expressions for the end-to-end SNDR statistics for the multibeam VHTS system with FSO feeder links, accounting for nonlinearities at satellite transponder.
\subsection{Cumulative Distribution Function}
A unified expression for the CDF of the overall SNDR at the $i$-th UT considering both IM/DD and heterodyne detection techniques for the FSO feeder link in the presence of HPA nonlinearity can be derived in exact closed-form in terms of the bivariate Meijer's G function whose implementation is presented in \cite{ImranBER,BivariateGCode} as
\begin{align}\label{CDFtotal}
\nonumber  F_{\gamma_i}(x)&= 1-\frac{\xi^2\, r^{\alpha+\beta-2}}{\Gamma(\alpha)\Gamma(\beta) (2 \pi)^{r-1}}\left (\frac{2b_i m_i}{2 b_i m_i+\Omega_i}  \right )^{m_i-1} \\
\nonumber & \times \sum_{k=0}^{m_i-1}\sum_{j=0}^{k}\frac{(-1)^k (1-m_i)_k}{k!\,j!}\left (\frac{\Omega_i}{2 b_i m_i}  \right )^k \\ & \times {\rm{G}}_{1,0:0,2:3r,r+1}^{1,0:2,0:0,3r}
\begin{bmatrix}
\begin{matrix}
0
\end{matrix}
\Bigg|\begin{matrix}
-\\j,1
\end{matrix}
\Bigg|\begin{matrix}
\mathcal{K}_1\\\Delta (r,-\xi^2),0
\end{matrix}
\Bigg|
\frac{C m_i}{\kappa\,\overline{\gamma}_{2,i}},\frac{r^{2r}(\xi^2+1)^r \mu_r }{(\alpha \beta \xi^2)^r \kappa \left \| \bb_i^{\mbox{\tiny T}} \right \|^2  x}
\end{bmatrix},
\end{align}
where $C=\tr\left [ \left(\bB \bB^{\mbox{\tiny H}} \right)^{-1} \right ]\overline{\gamma}_1+\kappa$, $\mathcal{K}_1=\Delta (r,1-\xi^2),\Delta (r,1-\alpha),\Delta (r,1-\beta)$, and $\Delta (r,u)=\frac{u}{r},\frac{u+1}{r},\ldots,\frac{u+r-1}{r}$.
\begin{IEEEproof}
See Appendix \ref{appendix:A}.
\end{IEEEproof}\noindent
Note that by setting $\kappa=1$ in (\ref{CDFtotal}), we can easily obtain the CDF expression in the case of linear PA at the satellite transponder.

\subsection{Probability Distribution Function}
The PDF of the end-to-end SNDR at the $i$-th UT, can be obtained by taking the derivative of (\ref{CDFtotal}), yielding
\begin{align}\label{PDFtotal}
\nonumber  f_{\gamma_i}(x)&= \frac{\xi^2\, r^{\alpha+\beta-2}}{\Gamma(\alpha)\Gamma(\beta) (2 \pi)^{r-1}x}\left (\frac{2b_i m_i}{2 b_i m_i+\Omega_i}  \right )^{m_i-1} \\
\nonumber & \times \sum_{k=0}^{m_i-1}\sum_{j=0}^{k}\frac{(-1)^k (1-m_i)_k}{k!\,j!}\left (\frac{\Omega_i}{2 b_i m_i}  \right )^k \\
&\times {\rm{G}}_{1,0:0,2:3r,r+1}^{1,0:2,0:0,3r}\begin{bmatrix}
\begin{matrix}
0
\end{matrix}
\Bigg|\begin{matrix}
-\\j,1
\end{matrix}
\Bigg|\begin{matrix}
\mathcal{K}_1\\\Delta (r,-\xi^2),1
\end{matrix}
\Bigg|
\frac{C m_i}{\kappa\,\overline{\gamma}_{2,i}},\frac{r^{2r}(\xi^2+1)^r \mu_r }{(\alpha \beta \xi^2)^r \kappa \left \| \bb_i^{\mbox{\tiny T}} \right \|^2 x}
\end{bmatrix},
\end{align}
\subsection{Moments}
The $n$-th moments of the end-to-end SNDR at the $i$-th UT defined as $\mathbb{E}[\gamma_i^n]\triangleq \int_{0}^{\infty}x^n\,f_{\gamma,i}(x)\,dx
$, can be given as
\begin{align}\label{moments}
\nonumber &\mathbb{E}[\gamma_i^n]=\frac{\xi^2\Gamma(\alpha+r\,n)\Gamma(\beta+ r\,n)}{(\xi^2+r\,n)\Gamma(\alpha)\Gamma(\beta)\Gamma(n)}\left (\frac{(\xi^2+1)^r \mu_r}{(\alpha \beta \xi^2)^r \kappa  \left \| \bb_i^{\mbox{\tiny T}} \right \|^2}  \right )^n\\
& \times \left (\frac{2b_i m_i}{2 b_i m_i+\Omega_i}  \right )^{m_i-1}
\sum_{k=0}^{m_i-1}\sum_{j=0}^{k}\frac{(-1)^k (1-m_i)_k}{k!\,j!}\left (\frac{\Omega_i}{2 b_i m_i}  \right )^k
{\rm{G}}_{1,2}^{2,1}\left[\frac{C m_i}{\kappa \overline{\gamma}_{2,i}}\left| \begin{matrix} {1-n} \\ {j,1} \\ \end{matrix} \right. \right].
\end{align}

\begin{IEEEproof}
See Appendix \ref{appendix:B}.
\end{IEEEproof}

\section{Performance Evaluation}
This section derives new closed-form expressions for the performance metrics of the multibeam VHTS system with an FSO feeder link
under the presence of satellite transponder nonlinearity. Additionally, this section provides tractable asymptotic expressions for the outage probability and the average BER at the high SNR regime.
\subsection{Outage Probability}
\subsubsection{Exact Analysis}
The outage probability is defined as the probability that the end-to-end SNDR falls below a predefined threshold $\gamma_{\rm{th}}$ and can be easily obtained at the $i$-th UT by setting $x=\gamma_{\rm{th}}$ in (\ref{CDFtotal}), that is,
\begin{align}\label{OP}
P_{{\mathrm{out}},i}(\gamma_{th})=F_{\gamma_i}(\gamma_{th}).
\end{align}
It can be concluded from (\ref{OP}) that for low values of IBO, the term $\kappa$ grows very large and the outage probability $P_{{\mathrm{out}},i}(\gamma_{th}) \to 1$ for any $\gamma_{th}$, especially in the case of TWTA. This shows the deleterious impact of the nonlinear amplifier at the relay and demonstrates that it is necessary to increase IBO in order to reduce the distortion introduced by both TWTA and SSPA models. To obtain more engineering insights on the impact of the hardware impairments, we elaborate further on the asymptotic analysis at high SNR regime.
\subsubsection{Asymptotic Analysis}
Starting from (\ref{CDFP3}), applying \cite[Eq.(1.5.9)]{HTranforms} then \cite[Eq.(1.8.4)]{HTranforms}, the outage probability at the $i$-th UT can be given asymptotically at high SNR of the FSO link after performing some algebraic manipulations as
\begin{align}\label{asymOP}
\nonumber  P_{{\mathrm{out}},i}(\gamma_{\rm{th}})& \underset{\mu_r\gg 1}{\mathop{\approx }}
1-\frac{\xi^2}{\Gamma(\alpha)\Gamma(\beta)}\left (\frac{2b_i m_i}{2 b_i m_i+\Omega_i}  \right )^{m_i-1} \\
& \times \sum_{k=0}^{m_i-1}\sum_{j=0}^{k}\frac{(-1)^k (1-m_i)_k}{k!\,j!}\left (\frac{\Omega_i}{2 b_i m_i}  \right )^k\sum_{v=1}^{4} \mathcal{J}_v \, \left(\frac{\gamma_{\rm{th}}}{\mu_r}\right)^{\theta_v},
\end{align}
where $\theta_v=\left \{ j, \frac{\xi^2}{r} ,\frac{\alpha}{r},\frac{\beta}{r}  \right \}$ and
\begin{align}
\mathcal{J}_1=\frac{\Gamma(\alpha-rj)\Gamma(\beta-rj)}{\xi^2-rj}\left (\frac{C\, m_i\,\left \| \bb_i^{\mbox{\tiny T}} \right \|^2(\alpha \beta \xi^2)^r }{(\xi^2+1)^r\,  \overline{\gamma}_{2,i}}\right )^j,
\end{align}\noindent
\begin{align}
\nonumber \mathcal{J}_2&=\frac{\Gamma(\alpha-\xi^2)\Gamma(\beta-\xi^2)}{r}\left (\frac{\kappa \left \| \bb_i^{\mbox{\tiny T}} \right \|^2(\alpha \beta \xi^2)^r }{(\xi^2+1)^r\, }\right )^{\frac{\xi^2}{r}}\\
& \times\left(\Gamma\left(j-\frac{\xi^{2}}{r}\right)\left(\frac{C m_i }{\kappa \overline{\gamma}_{2,i}}\right )^{\frac{\xi^{2}}{r}}+\frac{{\rm{G}}_{1,2}^{2,1}\left[\frac{C m_i}{\kappa \overline{\gamma}_{2,i}}\left| \begin{matrix} {1+\frac{\xi^2}{r}} \\ {j,1} \\ \end{matrix} \right. \right]}{\Gamma\left ( 1- \frac{\xi^2}{r}\right )}\right ),
\end{align}\noindent
\begin{align}
\nonumber \mathcal{J}_3&=\frac{\Gamma(\beta-\alpha)}{r(\xi^2-\alpha)}\left (\frac{\kappa \left \| \bb_i^{\mbox{\tiny T}} \right \|^2(\alpha \beta \xi^2)^r }{(\xi^2+1)^r\, }\right )^{\frac{\alpha}{r}}\\
& \times \left ( \Gamma\left ( j- \frac{\alpha}{r}
\right ) \left (\frac{C m_i }{\kappa \overline{\gamma}_{2,i}}\right )^\frac{\alpha}{r}+\frac{{\rm{G}}_{1,2}^{2,1}\left[\frac{C m_i}{\kappa \overline{\gamma}_{2,i}}\left| \begin{matrix} {1+\frac{\alpha}{r}} \\ {j,1} \\ \end{matrix} \right. \right]}{\Gamma\left ( 1- \frac{\alpha}{r}\right )}\right ),
\end{align}\noindent
\begin{align}
\nonumber \mathcal{J}_4&=\frac{\Gamma(\alpha-\beta)}{r(\xi^2-\beta)}\left (\frac{\kappa \left \| \bb_i^{\mbox{\tiny T}} \right \|^2(\alpha \beta \xi^2)^r }{(\xi^2+1)^r\, }\right )^{\frac{\beta}{r}}\\
& \times \left ( \Gamma\left ( j- \frac{\beta}{r}
\right ) \left (\frac{C m_i }{\kappa \overline{\gamma}_{2,i}}\right )^\frac{\beta}{r}+\frac{{\rm{G}}_{1,2}^{2,1}\left[\frac{C m_i}{\kappa \overline{\gamma}_{2,i}}\left| \begin{matrix} {1+\frac{\beta}{r}} \\ {j,1} \\ \end{matrix} \right. \right]}{\Gamma\left ( 1- \frac{\beta}{r}\right )}\right ).
\end{align}\noindent
Note that at high SNR of the RF user link, $\overline{\gamma}_{2,i}$, $\mathcal{J}_2$, $\mathcal{J}_3$ , and $\mathcal{J}_4$ can be further simplified by using
\cite[Eq.(B.1)]{HARQFSO} as
\begin{align}
 \mathcal{J}_2=\frac{2\Gamma(\alpha-\xi^2)\Gamma(\beta-\xi^2)\Gamma\left(j-\frac{\xi^2}{r}\right)}{r}\left (\frac{C m_i \left \| \bb_i^{\mbox{\tiny T}} \right \|^2(\alpha \beta \xi^2)^r }{(\xi^2+1)^r\,  \overline{\gamma}_{2,i}}\right )^{\frac{\xi^2}{r}},
\end{align}
\begin{align}
\mathcal{J}_3=\frac{2\Gamma(\beta-\alpha)\Gamma(j-\frac{\alpha}{r})}{r(\xi^2-\alpha)}\left (\frac{C\, m_i\,\left \| \bb_i^{\mbox{\tiny T}} \right \|^2(\alpha \beta \xi^2)^r }{(\xi^2+1)^r\,  \overline{\gamma}_{2,i}}\right )^{\frac{\alpha}{r}},
\end{align}\noindent
\begin{align}
\mathcal{J}_4=\frac{2\Gamma(\alpha-\beta)\Gamma(j-\frac{\beta}{r})}{r(\xi^2-\beta)}\left (\frac{C\, m_i\,\left \| \bb_i^{\mbox{\tiny T}} \right \|^2(\alpha \beta \xi^2)^r }{(\xi^2+1)^r\,  \overline{\gamma}_{2,i}}\right )^{\frac{\beta}{r}}.
\end{align}

It can be inferred from (\ref{asymOP}) that when $\mu_r \to \infty$, the parameter $\kappa$ grows towards infinity too and the outage probability is saturated by an irreducible floor regardless of the nonlinear HPA model at the relay. Indeed, the impairments become very severe at high SNR range and the outage probability does not decrease with an increase in the average electrical SNR. However, in the case of linear PA at the relay, the outage probability converges to zero when $\mu_r \to \infty$. This confirms that the hardware impairments can significantly
limit the performance of VHTS systems and therefore should be considered in the design of such systems.
\subsection{Average BER}
\subsubsection{Exact Analysis}
A generalized expression for the average BER of the $i$-th UT for a variety of modulation schemes can be expressed as  \cite[Eq.(22)]{dualhopFSO}
\begin{align}\label{BERDEF}
\overline{P}_{e,i}= \frac{\delta }{2\Gamma(p)}\sum_{u=1}^{n}q_u^p\int_{0}^{\infty}x^{p-1}e^{-q_u x}F_{\gamma_i}(x)\,dx,
\end{align}
where $n$, $\delta$, $p$, and $q_u$ vary depending on the modulation technique and the type of detection (i.e IM/DD or heterodyne detection) and are listed in Table~\ref{modulation}.
 It is important to mention here that for IM/DD technique, we investigate the average BER for on-off keying (OOK) modulation since it is the most commonly used intensity modulation technique in practical FSO systems due to its simplicity and resilience to laser nonlinearity. For heterodyne detection and in addition to binary modulation schemes, we analyze the average BER for multilevel phase shift keying (MPSK) and quadrature amplitude (MQAM) that are commonly deployed in coherent systems.
By substituting (\ref{CDFP4}) in (\ref{BERDEF}), integrating
using \cite[Eq.(3.381/4)]{Tableofintegrals} and applying \cite[Eq.(1)]{Gupta}, a unified expression for the average BER of the $i$-th UT for all these modulation schemes can be derived in exact closed-form in terms of the bivariate Meijer's G function as
\begin{align}\label{BERtotal}
\nonumber \overline{P}_{e,i}&= \frac{\delta\, n}{2}-\frac{\delta\,\xi^2\, r^{\alpha+\beta-2}}{2\Gamma(\alpha)\Gamma(\beta)\Gamma(p) (2 \pi)^{r-1}}\left (\frac{2b_i m_i}{2 b_i m_i+\Omega_i}  \right )^{m_i-1} \\
\nonumber & \times \sum_{k=0}^{m_i-1}\sum_{j=0}^{k}\sum_{u=1}^{n}\frac{(-1)^k (1-m_i)_k}{k!\,j!}\left (\frac{\Omega_i}{2 b_i m_i}  \right )^k \\
& \times {\rm{G}}_{1,0:0,2:3r,r+2}^{1,0:2,0:1,3r}
\begin{bmatrix}
\begin{matrix}
0
\end{matrix}
\Bigg|\begin{matrix}
-\\j,1
\end{matrix}
\Bigg|\begin{matrix}
\mathcal{K}_1\\p, \Delta (r,-\xi^2),0
\end{matrix}
\Bigg|
\frac{C m_i}{\kappa\,\overline{\gamma}_{2,i}},\frac{r^{2r}q_u(\xi^2+1)^r \mu_r }{(\alpha \beta \xi^2)^r   \kappa \left \| \bb_i^{\mbox{\tiny T}} \right \|^2}
\end{bmatrix}.
\end{align}
\begin{table}[!h]
\begin{minipage}{1\textwidth}
\normalsize
\begin{center}
\caption{Parameters for Different Modulation Schemes}\label{modulation}
\small{\begin{tabular}{@{}llllll@{}}
\toprule
\multicolumn{1}{c}{Modulation} & \multicolumn{1}{c}{${\delta}$} & \multicolumn{1}{c}{${p}$} & \multicolumn{1}{c}{${q_u}$} & \multicolumn{1}{c}{${n}$} & \multicolumn{1}{c}{{Type of Detection}} \\ \midrule
{OOK}        & $1$                         & $1/2$    & $1/2$      & $1$      & IM/DD                   \\
{BPSK}       & $1$                         & $1/2$    & $1$        & $1$      & Heterodyne              \\
{M-PSK}      & $\frac{2}{\max(\log_2M,2)}$ & $1/2$    &     $\sin^2\left ( \frac{(2u-1)\pi} {M} \right )$       & $\max\left(\frac{M}{4},1\right)$         &          Heterodyne               \\
{M-QAM}      & $\frac{4}{\log_2M}\left ( 1-\frac{1}{\sqrt{M}} \right )$  &   $1/2$       & $\frac{3(2u-1)^2}{2(M-1)}$  & $\frac{\sqrt{M}}{2}$         & Heterodyne                        \\ \bottomrule
\end{tabular}}
\end{center}
\end{minipage}
\end{table}\noindent \normalsize

It is worth noting that the major advantage of (\ref{BERtotal}) is that it presents a unified BER expression in a compact form that is valid
for both heterodyne and IM/DD techniques and applicable to a variety of modulation schemes. In addition, to reveal some useful insights, we derive an asymptotic expression for the average BER at high SNR regime as shown by (\ref{asymBER}).
\subsubsection{Asymptotic Analysis}
Similar to the asymptotic outage probability analysis, a simpler closed-form expression for the average BER of the $i$-th UT for a variety of modulation techniques can be obtained at high SNR regime by substituting the CDF expression at high SNR, obtained from (\ref{asymOP}), into (\ref{BERDEF}), and utilizing \cite[Eq.(3.381/4)]{Tableofintegrals} as

\begin{align}\label{asymBER}
\nonumber  \overline{P}_{e,i}& \underset{\mu_r\gg 1}{\mathop{\approx }} \frac{\delta\, n}{2}-\frac{\delta\,\xi^2}{2\Gamma(\alpha)\Gamma(\beta)\Gamma(p) }\left (\frac{2b_i m_i}{2 b_i m_i+\Omega_i}  \right )^{m_i-1}\sum_{k=0}^{m_i-1}\sum_{j=0}^{k} \\
 & \times \sum_{u=1}^{n}\frac{(-1)^k (1-m_i)_k}{k!\,j!}\left (\frac{\Omega_i}{2 b_i m_i}  \right )^k \sum_{v=1}^{4} \mathcal{J}_v \Gamma(p+\theta_v)(q_u \mu_r)^{-\theta_v}.
\end{align}

It is important to mention that the asymptotic expression of the average BER given in (\ref{asymBER}) is simpler and much more analytically tractable than the exact expression of the BER obtained in terms of the bivariate Meijer's G function in (\ref{BERtotal}),
which is a quite complex function and not a standard built-in function in most of the well-known
mathematical software tools such as MATHEMATICA and MATLAB. Interestingly, (\ref{asymBER}) is
very accurate and converges perfectly to the exact result in (\ref{BERtotal}) at high SNR regime, which
is illustrated in section V. Similar to what was concluded from the asymptotic expression of the outage probability in (\ref{asymOP}), it can be easily shown that a BER floor is created at high SNR range due to HPA nonlinearity, which becomes higher as IBO gets lower.
\subsection{Ergodic Capacity}
\subsubsection*{Exact Analysis}
The ergodic capacity of the $i$-th UT of an FSO-based mutilbeam VHTS system with HPA nonlinearity can be calculated as \cite{Anas2016},\cite[Eq.(26)]{lapidoth} \cite[Eq.(7.43)]{owc}
\begin{align}\label{CAPDEF}
\overline{C_i}\triangleq \mathbb{E}[\ln(1+\tau\,\gamma_i)]=\frac{\tau}{\ln(2)}\int_{0}^{\infty}(1+\tau\,x)^{-1}F_{\gamma_i}^c(x)\,dx,
\end{align}
where $\tau=e/(2\pi)$ for the IM/DD technique and $\tau=1$ for the heterodyne detection technique. It is worthy to mention that the expression in (\ref{CAPDEF}) is exact for $r=1$ while it is a lower-bound for $r=2$, and can be achieved in exact closed-form in terms of the bivariate Meijer's G function by substituting (\ref{CDFP4}) into (\ref{CAPDEF}), applying \cite[Eqs.(07.34.03.0271.01) and (07.34.21.0009.011)]{Wolfram}, and using \cite[Eq.(1)]{Gupta} as
\begin{align}\label{Capacitytotal}
\nonumber  \overline{C}_{i}& = \frac{\xi^2\, r^{\alpha+\beta-2}}{\ln (2)\Gamma(\alpha)\Gamma(\beta) (2 \pi)^{r-1}}\left (\frac{2b_i m_i}{2 b_i m_i+\Omega_i}  \right )^{m_i-1} \\
\nonumber & \times \sum_{k=0}^{m_i-1}\sum_{j=0}^{k}\frac{(-1)^k (1-m_i)_k}{k!\,j!}\left (\frac{\Omega_i}{2 b_i m_i}  \right )^k \\
& \times {\rm{G}}_{1,0:0,2:3r+1,r+2}^{1,0:2,0:1,3r+1}
\begin{bmatrix}
\begin{matrix}
0
\end{matrix}
\Bigg|\begin{matrix}
-\\j,1
\end{matrix}
\Bigg|\begin{matrix}
1,\mathcal{K}_1\\1,\Delta (r,-\xi^2),0
\end{matrix}
\Bigg|
\frac{C m_i}{\kappa\,\overline{\gamma}_{2,i}},\frac{r^{2r}\tau(\xi^2+1)^r \mu_r }{(\alpha \beta \xi^2)^r \kappa \left \| \bb_i^{\mbox{\tiny T}} \right \|^2}
\end{bmatrix},
\end{align}
In the special case when $\kappa=1$, (50) reduces to the ergodic capacity of the $i$-th UT of an FSO-based mutilbeam VHTS system with linear PA at the satellite transponder.
\section{Numerical Results}
In this section, we examine the performance of a FSO-based multibeam VHTS system in the presence of atmospheric turbulence, beam wander effect, pointing errors, and HPA nonlinearities using the set of parameters listed in Table~\ref{table:sysparam} \cite{SatelliteFSO,Andrews}.
\begin{table}[!h]
\begin{center}
\caption{System Parameters}\label{table:sysparam}
\begin{tabular}{l l l}
    \toprule
    Parameter & Value\\[0.5ex]
    \midrule
    Altitude of the satellite, $H$ & $35786 \times 10^3$ m  (GEO)\\[1ex]
    Altitude of the OGS, $h_0$& 0 m\\[1ex]
    Zenith angle, $\zeta$& $30^{o}$\\[1ex]
Carrier frequency, $f$ & 20 Ghz (Ka-band) \\[1ex]
    Number of beams, $N$ & 7 \\[1ex]
Beam radius, $R$ & 250 Km \\[1ex]
Satellite antenna gain, $G_r$ & 52 dBi \\[1ex]
UT antenna gain, $G_t$ & 38.16 dBi \\[1ex]
 Noise bandwidth, $B_w$ & 50 MHz \\[1ex]
 Boltzman constant, $\kappa_B$ & $1.38 \times 10^{-23}$ J/K \\[1ex]
Receiver noise temperature, $T_r$ & 207 K \\[1ex]
3 dB angle, $\theta_{3\rm{dB}}$ & $0.4^{o}$\\[1ex]
Transmitter beam radius, $W_0$ & 0.02 m\\[1ex]
Wind velocity, $w$ & 21m/s\\[1ex]
Wavelength, $\lambda$ & $1550$ nm\\[1ex]
Phase front radius of curvature, $F_0$  & $\infty$\\[1ex]
         \bottomrule
    \end{tabular}
\end{center}
\end{table}\noindent
Monte-Carlo simulations are also included and compared with the obtained analytical results over $10^6$ realizations. A very good match between all the derived and the respective simulated results is observed, and hence, the accuracy of the proposed framework is verified.
As illustrated in Fig.~\ref{fig:systemmodel}, we consider a FSS system with a coverage area composed of $N=7$ beams that serves multiple single antenna UTs. Assuming that there is only one UT per beam, each UT has a fixed position within each beam as shown by Fig.~\ref{fig:systemmodel}.
Without loss of generality, we consider three types of turbulence conditions based on three values of nominal ground turbulence levels, i.e. $C_n^2(0)=1\times 10^{-13} {\rm m}^{-\frac{2}{3}}$, $C_n^2(0)=5\times 10^{-13} {\rm m}^{-\frac{2}{3}}$, and $C_n^2(0)=1\times 10^{-12} {\rm m}^{-\frac{2}{3}}$ \cite{Subrat2015}.
Hence, from (\ref{alphaEq}) and (\ref{betaEq}), the scintillation parameters ($\alpha$, $\beta$) can be computed as (8.41, 14.67) with $\sigma_{\rm{pe}}=154.9$, (2.57, 5.36) with $\sigma_{\rm{pe}}=141.59$, and (1.52, 3.29) with $\sigma_{\rm{pe}}=133.18$, respectively, when beam wander effects are included, whereas ($\alpha$, $\beta$)=(15.4, 14.67), (5.76, 5.36), and (3.62, 3.29), respectively, when beam wander effects are ignored.
In addition, for the RF user link, two channel fading conditions are considered, namely, unfrequent light $\left \{ m_i, b_i, \Omega_i \right \}=\left \{ 19, 0.158, 1.29 \right \}$ and frequent heavy $\left \{ m_i, b_i, \Omega_i \right \}=\left \{ 1, 0.063, 8.97\times 10^{-4} \right \}$ shadowing as provided in \cite[Table III]{LMSAlouini}. Moreover, Since the noise power is normalized by $\kappa_B B_w T_r$ in (\ref{Bij}), we can assume that $\sigma_1^2=1$ and we select $\sigma_2^2=1$, $I_l=1$, and $G=1$. Furthermore, we evaluate the performance relevant to the central beam which is located in the center of the coverage area and receives the maximum interference from the six adjacent beams as illustrated in Fig.~\ref{fig:systemmodel}.

The outage probability versus the average electrical SNR $\mu_r$ of the FSO feeder link under SSPA and TWTA models is plotted in Fig.~\ref{fig:OP1} for different values of $\gamma_{\rm{th}}$. For both nonlinear HPA models, IBO is set to 25 dB. Results of the linear PA are also included for comparison purposes. It can be observed that the outage performance improves with the increase of $\mu_r$ up to 35 dB under both TWTA and SSPA models. Moreover, both TWTA and SSPA as well as linear PA have the same impact on the outage probability up to 35 dB of the average electrical SNR. However, when $\mu_r$ exceeds 35 dB, the nonlinearity effect of the power amplifier becomes more pronounced and the outage probability does not decrease even if $\mu_r$ proceeds to increase. Indeed, as $\mu_r$ gets larger, an outage floor is introduced regardless of the nonlinear HPA model while, it does not occur in the case of linear PA system that evidently performs better than the system with nonlinear power amplifier. Also, it can be noted that above 35 dB, TWTA and SSPA have different effects on the outage performance and the degradation of the outage probability caused by TWTA model is the largest.
Furthermore, it is evident that the greater the value of the effective SNDR $\gamma_{\rm{th}}$, the higher will be the outage probability of the system for both HPA models.
The asymptotic results of the outage probability at high average electrical SNR values obtained by using (\ref{asymOP}) are also included in Fig.~\ref{fig:OP1}. Obviously, the asymptotic results of the outage probability match perfectly the analytical results in the high SNR regime. This justifies the accuracy and the tightness of the derived asymptotic expression in (\ref{asymOP}).
\begin{figure}[!h]
  \begin{center}
\begin{tikzpicture}[scale=1.2]
    \begin{axis}[font=\footnotesize,
    ymode=log, xlabel= Average Electrical SNR $\mu_r$ (dB), ylabel= Outage Probability,
  xmin=0,xmax=80,ymax=1,ymin=1e-04,
    legend style={nodes=right},legend pos= south west,legend style={nodes={scale=1, transform shape}},
      xminorgrids,
    grid style={dotted},
    yminorgrids,
   ]
     \addplot[smooth,blue,mark=triangle,mark options={solid},every mark/.append style={solid, fill=white}] plot coordinates {
(0.000000,9.917999e-01)(5.000000,9.587234e-01)(10.000000,8.741945e-01)(15.000000,7.321990e-01)(20.000000,5.592319e-01)(25.000000,3.941946e-01)(30.000000,2.645801e-01)(35.000000,1.793484e-01)(40.000000,1.331555e-01)(45.000000,1.129409e-01)(50.000000,1.055055e-01)(55.000000,1.030068e-01)(60.000000,1.021958e-01)(65.000000,1.019422e-01)(70.000000,1.018611e-01)(75.000000,1.018345e-01)(80.000000,1.018274e-01)
};
    \addplot[smooth,green!70!black,mark=*,mark options={solid},every mark/.append style={solid, fill=white}] plot coordinates {
(0.000000,9.917529e-01)(5.000000,9.587148e-01)(10.000000,8.739373e-01)(15.000000,7.318234e-01)(20.000000,5.577126e-01)(25.000000,3.899034e-01)(30.000000,2.549882e-01)(35.000000,1.605884e-01)(40.000000,1.025877e-01)(45.000000,7.206260e-02)(50.000000,5.867770e-02)(55.000000,5.365520e-02)(60.000000,5.194810e-02)(65.000000,5.138870e-02)(70.000000,5.120970e-02)(75.000000,5.115740e-02)(80.000000,5.114020e-02)
};
    \addplot[smooth,cyan,mark=diamond,mark options={solid},every mark/.append style={solid, fill=white}] plot coordinates {
(0.000000,9.917621e-01)(5.000000,9.588028e-01)(10.000000,8.738812e-01)(15.000000,7.315238e-01)(20.000000,5.570199e-01)(25.000000,3.882607e-01)(30.000000,2.513555e-01)(35.000000,1.533023e-01)(40.000000,8.923270e-02)(45.000000,5.016360e-02)(50.000000,2.752600e-02)(55.000000,1.474100e-02)(60.000000,7.811300e-03)(65.000000,4.084100e-03)(70.000000,2.129000e-03)(75.000000,1.098000e-03)(80.000000,5.646000e-04)
};
     \addplot[smooth,blue,mark=triangle,mark options={solid},every mark/.append style={solid, fill=white}] plot coordinates {
(0.000000,5.571391e-01)(5.000000,3.888652e-01)(10.000000,2.518382e-01)(15.000000,1.537005e-01)(20.000000,9.004620e-02)(25.000000,5.133620e-02)(30.000000,2.941340e-02)(35.000000,1.792090e-02)(40.000000,1.245990e-02)(45.000000,1.024240e-02)(50.000000,9.468600e-03)(55.000000,9.201200e-03)(60.000000,9.111100e-03)(65.000000,9.085000e-03)(70.000000,9.076100e-03)(75.000000,9.073100e-03)(80.000000,9.072500e-03)
};
    \addplot[smooth,green!70!black,mark=*,mark options={solid},every mark/.append style={solid, fill=white}] plot coordinates {
(0.000000,5.572728e-01)(5.000000,3.888421e-01)(10.000000,2.517042e-01)(15.000000,1.537006e-01)(20.000000,8.969800e-02)(25.000000,5.064130e-02)(30.000000,2.806220e-02)(35.000000,1.566860e-02)(40.000000,9.225800e-03)(45.000000,6.152200e-03)(50.000000,4.889200e-03)(55.000000,4.429400e-03)(60.000000,4.280500e-03)(65.000000,4.232300e-03)(70.000000,4.214500e-03)(75.000000,4.210200e-03)(80.000000,4.208500e-03)
};
    \addplot[smooth,cyan,mark=diamond,mark options={solid},every mark/.append style={solid, fill=white}] plot coordinates {
(0.000000,5.572381e-01)(5.000000,3.887251e-01)(10.000000,2.517331e-01)(15.000000,1.535306e-01)(20.000000,8.950000e-02)(25.000000,5.035180e-02)(30.000000,2.760140e-02)(35.000000,1.482190e-02)(40.000000,7.801500e-03)(45.000000,4.090700e-03)(50.000000,2.119600e-03)(55.000000,1.082900e-03)(60.000000,5.536000e-04)(65.000000,2.825000e-04)(70.000000,1.482000e-04)(75.000000,7.580000e-05)(80.000000,4.010000e-05)
};
     \addplot[smooth,red,densely dashed,  thick] plot coordinates {
(0.000000,-1.974280e+03)(5.000000,-4.116020e+02)(10.000000,-7.771840e+01)(15.000000,-1.456290e+01)(20.000000,-2.597870e+00)(25.000000,-2.893030e-01)(30.000000,1.200410e-01)(35.000000,1.549050e-01)(40.000000,1.328790e-01)(45.000000,1.150040e-01)(50.000000,1.057850e-01)(55.000000,1.014550e-01)(60.000000,9.941590e-02)(65.000000,9.843520e-02)(70.000000,9.795430e-02)(75.000000,9.771560e-02)(80.000000,9.759630e-02)
};
     \addplot[smooth,red,densely dashed,  thick] plot coordinates {
(0.000000,-1.988540e+03)(5.000000,-4.122470e+02)(10.000000,-7.759410e+01)(15.000000,-1.448030e+01)(20.000000,-2.559900e+00)(25.000000,-2.739280e-01)(30.000000,1.212260e-01)(35.000000,1.429100e-01)(40.000000,1.072260e-01)(45.000000,7.823480e-02)(50.000000,6.277570e-02)(55.000000,5.585420e-02)(60.000000,5.290360e-02)(65.000000,5.162340e-02)(70.000000,5.104700e-02)(75.000000,5.077790e-02)(80.000000,5.064930e-02)
};
     \addplot[smooth,red,densely dashed,  thick] plot coordinates {
(0.000000,-2.002980e+03)(5.000000,-4.130500e+02)(10.000000,-7.755060e+01)(15.000000,-1.443790e+01)(20.000000,-2.541800e+00)(25.000000,-2.672730e-01)(30.000000,1.217840e-01)(35.000000,1.380640e-01)(40.000000,9.542870e-02)(45.000000,5.742160e-02)(50.000000,3.242190e-02)(55.000000,1.765970e-02)(60.000000,9.402250e-03)(65.000000,4.928810e-03)(70.000000,2.555290e-03)(75.000000,1.313960e-03)(80.000000,6.714700e-04)
};
     \addplot[smooth,red,densely dashed,  thick] plot coordinates {
(0.000000,-2.558670e+00)(5.000000,-2.722650e-01)(10.000000,1.205690e-01)(15.000000,1.380550e-01)(20.000000,9.596800e-02)(25.000000,5.850090e-02)(30.000000,3.430640e-02)(35.000000,2.074940e-02)(40.000000,1.402860e-02)(45.000000,1.108120e-02)(50.000000,9.887850e-03)(55.000000,9.410960e-03)(60.000000,9.214540e-03)(65.000000,9.130010e-03)(70.000000,9.092050e-03)(75.000000,9.074410e-03)(80.000000,9.066020e-03)
};
     \addplot[smooth,red,densely dashed,  thick] plot coordinates {
(0.000000,-2.550110e+00)(5.000000,-2.697190e-01)(10.000000,1.211780e-01)(15.000000,1.380080e-01)(20.000000,9.556980e-02)(25.000000,5.770820e-02)(30.000000,3.291630e-02)(35.000000,1.849170e-02)(40.000000,1.075030e-02)(45.000000,6.956240e-03)(50.000000,5.288890e-03)(55.000000,4.615880e-03)(60.000000,4.350550e-03)(65.000000,4.243380e-03)(70.000000,4.198210e-03)(75.000000,4.178270e-03)(80.000000,4.169170e-03)
};
     \addplot[smooth,red,densely dashed,  thick] plot coordinates {
(0.000000,-2.541800e+00)(5.000000,-2.672730e-01)(10.000000,1.217840e-01)(15.000000,1.380640e-01)(20.000000,9.542870e-02)(25.000000,5.742160e-02)(30.000000,3.242190e-02)(35.000000,1.765970e-02)(40.000000,9.402250e-03)(45.000000,4.928810e-03)(50.000000,2.555290e-03)(55.000000,1.313960e-03)(60.000000,6.714700e-04)(65.000000,3.414940e-04)(70.000000,1.730210e-04)(75.000000,8.739940e-05)(80.000000,4.404220e-05)
};
  \addplot[smooth, mark=star,only marks,red] plot coordinates {
(0.000000,9.917999e-01)(5.000000,9.587234e-01)(10.000000,8.741945e-01)(15.000000,7.321990e-01)(20.000000,5.592319e-01)(25.000000,3.941946e-01)(30.000000,2.645801e-01)(35.000000,1.793484e-01)(40.000000,1.331555e-01)(45.000000,1.129409e-01)(50.000000,1.055055e-01)(55.000000,1.030068e-01)(60.000000,1.021958e-01)(65.000000,1.019422e-01)(70.000000,1.018611e-01)(75.000000,1.018345e-01)(80.000000,1.018274e-01)
};
      \addplot[smooth, mark=star,only marks,red] plot coordinates {
(0.000000,9.917529e-01)(5.000000,9.587148e-01)(10.000000,8.739373e-01)(15.000000,7.318234e-01)(20.000000,5.577126e-01)(25.000000,3.899034e-01)(30.000000,2.549882e-01)(35.000000,1.605884e-01)(40.000000,1.025877e-01)(45.000000,7.206260e-02)(50.000000,5.867770e-02)(55.000000,5.365520e-02)(60.000000,5.194810e-02)(65.000000,5.138870e-02)(70.000000,5.120970e-02)(75.000000,5.115740e-02)(80.000000,5.114020e-02)
};
     \addplot[smooth, mark=star,only marks,red] plot coordinates {
(0.000000,9.917621e-01)(5.000000,9.588028e-01)(10.000000,8.738812e-01)(15.000000,7.315238e-01)(20.000000,5.570199e-01)(25.000000,3.882607e-01)(30.000000,2.513555e-01)(35.000000,1.533023e-01)(40.000000,8.923270e-02)(45.000000,5.016360e-02)(50.000000,2.752600e-02)(55.000000,1.474100e-02)(60.000000,7.811300e-03)(65.000000,4.084100e-03)(70.000000,2.129000e-03)(75.000000,1.098000e-03)(80.000000,5.646000e-04)
};
      \addplot[smooth, mark=star,only marks,red] plot coordinates {
(0.000000,5.571391e-01)(5.000000,3.888652e-01)(10.000000,2.518382e-01)(15.000000,1.537005e-01)(20.000000,9.004620e-02)(25.000000,5.133620e-02)(30.000000,2.941340e-02)(35.000000,1.792090e-02)(40.000000,1.245990e-02)(45.000000,1.024240e-02)(50.000000,9.468600e-03)(55.000000,9.201200e-03)(60.000000,9.111100e-03)(65.000000,9.085000e-03)(70.000000,9.076100e-03)(75.000000,9.073100e-03)(80.000000,9.072500e-03)
};
    \addplot[smooth, mark=star,only marks,red] plot coordinates {
(0.000000,5.572728e-01)(5.000000,3.888421e-01)(10.000000,2.517042e-01)(15.000000,1.537006e-01)(20.000000,8.969800e-02)(25.000000,5.064130e-02)(30.000000,2.806220e-02)(35.000000,1.566860e-02)(40.000000,9.225800e-03)(45.000000,6.152200e-03)(50.000000,4.889200e-03)(55.000000,4.429400e-03)(60.000000,4.280500e-03)(65.000000,4.232300e-03)(70.000000,4.214500e-03)(75.000000,4.210200e-03)(80.000000,4.208500e-03)
};
  \addplot[smooth, mark=star,only marks,red] plot coordinates {
(0.000000,5.572381e-01)(5.000000,3.887251e-01)(10.000000,2.517331e-01)(15.000000,1.535306e-01)(20.000000,8.950000e-02)(25.000000,5.035180e-02)(30.000000,2.760140e-02)(35.000000,1.482190e-02)(40.000000,7.801500e-03)(45.000000,4.090700e-03)(50.000000,2.119600e-03)(55.000000,1.082900e-03)(60.000000,5.536000e-04)(65.000000,2.825000e-04)(70.000000,1.482000e-04)(75.000000,7.580000e-05)(80.000000,4.010000e-05)
};

\legend{TWTA, SSPA, Linear PA, ,,,High SNR,,,,,,Simulation};

 \draw \boundellipse{ axis cs:47.5,7.206260e-02}{10}{1};
\node at (axis cs:47.5,3e-01){\scriptsize{$\gamma_{\rm{th}}=5$dB}};

\draw \boundellipse{ axis cs:47.5,6.152200e-03}{10}{1.3};
\node at (axis cs:45,0.9e-03){\scriptsize{$\gamma_{\rm{th}}=-15$dB}};

 \end{axis}
  \end{tikzpicture}
     \caption{OP under TWTA and SSPA models with 25 dB IBO for different values of $\gamma_{\rm{th}}$ when $\xi=1.1$ and $C_n^2(0)=1 \times 10^{-12}$ with beam wander effect under light shadowing conditions using IM/DD along with the asymptotic results at high SNR.}
          \label{fig:OP1}
     \end{center}
  \end{figure}
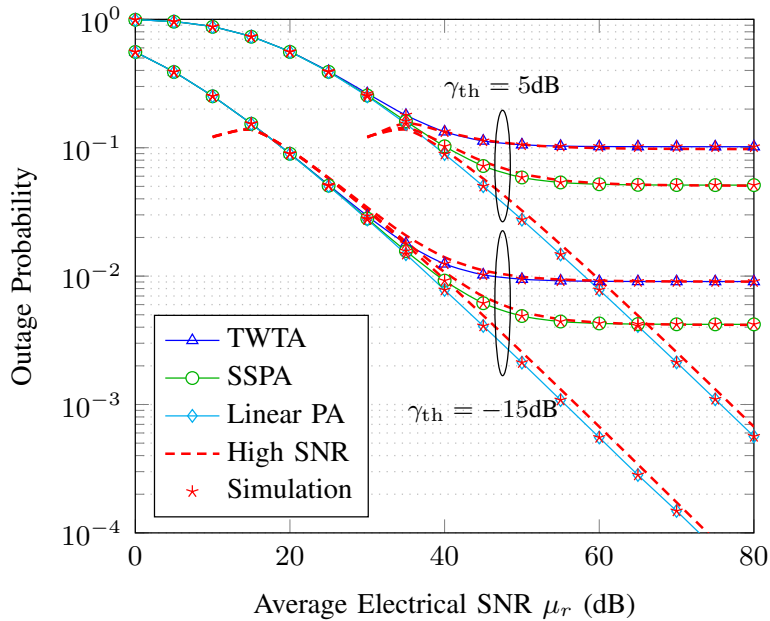\noindent

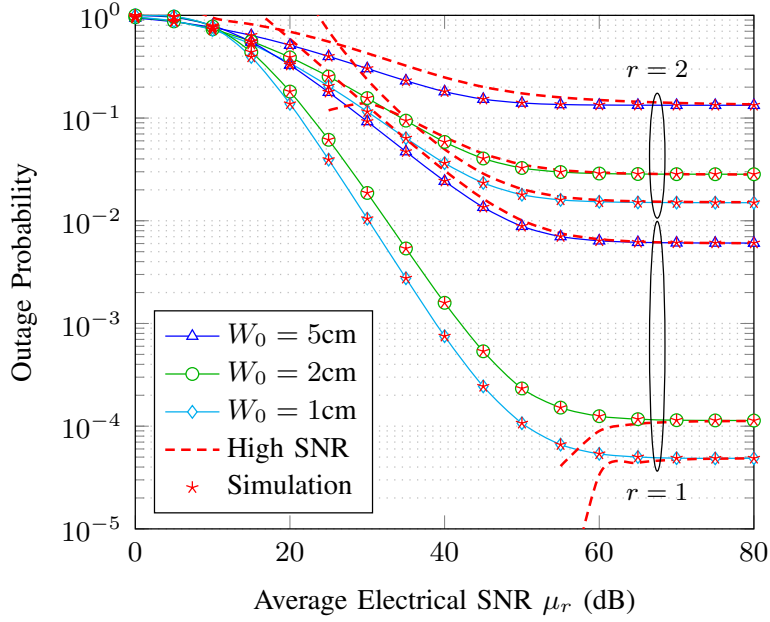
\begin{figure}[!h]
  \begin{center}
\begin{tikzpicture}[scale=1.2]
    \begin{axis}[font=\footnotesize,
    ymode=log, xlabel= Average Electrical SNR $\mu_r$ (dB), ylabel= Outage Probability,
  xmin=0,xmax=80,ymax=1,ymin=1e-05,
    legend style={nodes=right},legend pos= south west,legend style={nodes={scale=1, transform shape}},
      xminorgrids,
    grid style={dotted},
    yminorgrids,
   ]
     \addplot[smooth,blue,mark=triangle*,mark options={solid},every mark/.append style={solid, fill=white}] plot coordinates {
(0.000000,9.411214e-01)(5.000000,8.671436e-01)(10.000000,7.616509e-01)(15.000000,6.381314e-01)(20.000000,5.127244e-01)(25.000000,3.987979e-01)(30.000000,3.035513e-01)(35.000000,2.306339e-01)(40.000000,1.808859e-01)(45.000000,1.527927e-01)(50.000000,1.401985e-01)(55.000000,1.355188e-01)(60.000000,1.339555e-01)(65.000000,1.334424e-01)(70.000000,1.332772e-01)(75.000000,1.332231e-01)(80.000000,1.332062e-01)
};
    \addplot[smooth,green!70!black,mark=*,mark options={solid},every mark/.append style={solid, fill=white}] plot coordinates {
(0.000000,9.585219e-01)(5.000000,8.739215e-01)(10.000000,7.319189e-01)(15.000000,5.581028e-01)(20.000000,3.900437e-01)(25.000000,2.535883e-01)(30.000000,1.564737e-01)(35.000000,9.438140e-02)(40.000000,5.854980e-02)(45.000000,4.039390e-02)(50.000000,3.265770e-02)(55.000000,2.980000e-02)(60.000000,2.883630e-02)(65.000000,2.852460e-02)(70.000000,2.841970e-02)(75.000000,2.838990e-02)(80.000000,2.837950e-02)
};
    \addplot[smooth,cyan,mark=diamond*,mark options={solid},every mark/.append style={solid, fill=white}] plot coordinates {
(0.000000,9.667391e-01)(5.000000,8.797944e-01)(10.000000,7.207455e-01)(15.000000,5.242179e-01)(20.000000,3.418270e-01)(25.000000,2.043763e-01)(30.000000,1.153784e-01)(35.000000,6.369490e-02)(40.000000,3.639010e-02)(45.000000,2.336540e-02)(50.000000,1.794000e-02)(55.000000,1.597100e-02)(60.000000,1.531120e-02)(65.000000,1.509910e-02)(70.000000,1.502990e-02)(75.000000,1.500690e-02)(80.000000,1.499950e-02)
};
     \addplot[smooth,blue,mark=triangle*,mark options={solid},every mark/.append style={solid, fill=white}] plot coordinates {
(0.000000,9.976891e-01)(5.000000,9.561900e-01)(10.000000,7.961224e-01)(15.000000,5.501924e-01)(20.000000,3.280374e-01)(25.000000,1.785655e-01)(30.000000,9.254930e-02)(35.000000,4.709790e-02)(40.000000,2.431470e-02)(45.000000,1.354260e-02)(50.000000,8.841200e-03)(55.000000,7.023500e-03)(60.000000,6.386600e-03)(65.000000,6.169800e-03)(70.000000,6.097800e-03)(75.000000,6.074300e-03)(80.000000,6.067800e-03)
};
    \addplot[smooth,green!70!black,mark=*,mark options={solid},every mark/.append style={solid, fill=white}] plot coordinates {
(0.000000,9.996280e-01)(5.000000,9.725719e-01)(10.000000,7.793159e-01)(15.000000,4.385650e-01)(20.000000,1.810978e-01)(25.000000,6.144310e-02)(30.000000,1.865700e-02)(35.000000,5.371700e-03)(40.000000,1.589400e-03)(45.000000,5.358000e-04)(50.000000,2.330000e-04)(55.000000,1.516000e-04)(60.000000,1.251000e-04)(65.000000,1.170000e-04)(70.000000,1.143000e-04)(75.000000,1.138000e-04)(80.000000,1.134000e-04)
};
    \addplot[smooth,cyan,mark=diamond*,mark options={solid},every mark/.append style={solid, fill=white}] plot coordinates {
(0.000000,9.998746e-01)(5.000000,9.794049e-01)(10.000000,7.745260e-01)(15.000000,3.935506e-01)(20.000000,1.376127e-01)(25.000000,3.938780e-02)(30.000000,1.047220e-02)(35.000000,2.758700e-03)(40.000000,7.519000e-04)(45.000000,2.441000e-04)(50.000000,1.069000e-04)(55.000000,6.620000e-05)(60.000000,5.380000e-05)(65.000000,5.010000e-05)(70.000000,4.880000e-05)(75.000000,4.860000e-05)(80.000000,4.870000e-05)
};
     \addplot[smooth,red,densely dashed,  thick] plot coordinates {
(0.000000,5.025280e+00)(5.000000,1.473240e+00)(10.000000,9.498030e-01)(15.000000,8.191820e-01)(20.000000,6.928870e-01)(25.000000,5.548940e-01)(30.000000,4.273510e-01)(35.000000,3.236780e-01)(40.000000,2.486310e-01)(45.000000,2.010150e-01)(50.000000,1.739750e-01)(55.000000,1.588660e-01)(60.000000,1.498720e-01)(65.000000,1.441300e-01)(70.000000,1.402940e-01)(75.000000,1.376720e-01)(80.000000,1.358600e-01)
};
     \addplot[smooth,red,densely dashed,  thick] plot coordinates {
(0.000000,-3.118310e+02)(5.000000,-6.539990e+01)(10.000000,-1.308920e+01)(15.000000,-2.425560e+00)(20.000000,-2.673370e-01)(25.000000,1.188260e-01)(30.000000,1.389440e-01)(35.000000,9.955900e-02)(40.000000,6.530140e-02)(45.000000,4.506620e-02)(50.000000,3.532370e-02)(55.000000,3.117210e-02)(60.000000,2.946250e-02)(65.000000,2.874240e-02)(70.000000,2.842650e-02)(75.000000,2.828240e-02)(80.000000,2.821430e-02)
};
     \addplot[smooth,red,densely dashed,  thick] plot coordinates {
(0.000000,2.207530e+03)(5.000000,4.525840e+02)(10.000000,8.074760e+01)(15.000000,1.434500e+01)(20.000000,2.826200e+00)(25.000000,7.050440e-01)(30.000000,2.371540e-01)(35.000000,1.004340e-01)(40.000000,4.962900e-02)(45.000000,2.878310e-02)(50.000000,2.036740e-02)(55.000000,1.713760e-02)(60.000000,1.592150e-02)(65.000000,1.545420e-02)(70.000000,1.526720e-02)(75.000000,1.518860e-02)(80.000000,1.515430e-02)
};
     \addplot[smooth,red,densely dashed,  thick] plot coordinates {
(0.000000,3.224620e+02)(5.000000,1.146170e+01)(10.000000,2.541500e+00)(15.000000,1.223420e+00)(20.000000,5.893290e-01)(25.000000,2.783860e-01)(30.000000,1.312420e-01)(35.000000,6.258140e-02)(40.000000,3.079140e-02)(45.000000,1.634440e-02)(50.000000,1.008380e-02)(55.000000,7.565750e-03)(60.000000,6.612870e-03)(65.000000,6.257910e-03)(70.000000,6.122880e-03)(75.000000,6.069610e-03)(80.000000,6.047710e-03)
};
     \addplot[smooth,red,densely dashed,  thick] plot coordinates {
(55.000000,4.083800e-05)(60.000000,8.973350e-05)(65.000000,1.053220e-04)(70.000000,1.102570e-04)(75.000000,1.118170e-04)(80.000000,1.123090e-04)
};
     \addplot[smooth,red,densely dashed,  thick] plot coordinates {
(55.000000,1.053880e-06)(60.000000,3.357790e-05)(65.000000,4.393130e-05)(70.000000,4.720930e-05)(75.000000,4.824550e-05)(80.000000,4.857300e-05)

};
  \addplot[smooth, mark=star,only marks,red] plot coordinates {
(0.000000,9.411214e-01)(5.000000,8.671436e-01)(10.000000,7.616509e-01)(15.000000,6.381314e-01)(20.000000,5.127244e-01)(25.000000,3.987979e-01)(30.000000,3.035513e-01)(35.000000,2.306339e-01)(40.000000,1.808859e-01)(45.000000,1.527927e-01)(50.000000,1.401985e-01)(55.000000,1.355188e-01)(60.000000,1.339555e-01)(65.000000,1.334424e-01)(70.000000,1.332772e-01)(75.000000,1.332231e-01)(80.000000,1.332062e-01)
};
      \addplot[smooth, mark=star,only marks,red] plot coordinates {
(0.000000,9.585219e-01)(5.000000,8.739215e-01)(10.000000,7.319189e-01)(15.000000,5.581028e-01)(20.000000,3.900437e-01)(25.000000,2.535883e-01)(30.000000,1.564737e-01)(35.000000,9.438140e-02)(40.000000,5.854980e-02)(45.000000,4.039390e-02)(50.000000,3.265770e-02)(55.000000,2.980000e-02)(60.000000,2.883630e-02)(65.000000,2.852460e-02)(70.000000,2.841970e-02)(75.000000,2.838990e-02)(80.000000,2.837950e-02)
};
     \addplot[smooth, mark=star,only marks,red] plot coordinates {
(0.000000,9.667391e-01)(5.000000,8.797944e-01)(10.000000,7.207455e-01)(15.000000,5.242179e-01)(20.000000,3.418270e-01)(25.000000,2.043763e-01)(30.000000,1.153784e-01)(35.000000,6.369490e-02)(40.000000,3.639010e-02)(45.000000,2.336540e-02)(50.000000,1.794000e-02)(55.000000,1.597100e-02)(60.000000,1.531120e-02)(65.000000,1.509910e-02)(70.000000,1.502990e-02)(75.000000,1.500690e-02)(80.000000,1.499950e-02)
};
      \addplot[smooth, mark=star,only marks,red] plot coordinates {
(0.000000,9.976891e-01)(5.000000,9.561900e-01)(10.000000,7.961224e-01)(15.000000,5.501924e-01)(20.000000,3.280374e-01)(25.000000,1.785655e-01)(30.000000,9.254930e-02)(35.000000,4.709790e-02)(40.000000,2.431470e-02)(45.000000,1.354260e-02)(50.000000,8.841200e-03)(55.000000,7.023500e-03)(60.000000,6.386600e-03)(65.000000,6.169800e-03)(70.000000,6.097800e-03)(75.000000,6.074300e-03)(80.000000,6.067800e-03)
};
    \addplot[smooth, mark=star,only marks,red] plot coordinates {
(0.000000,9.996280e-01)(5.000000,9.725719e-01)(10.000000,7.793159e-01)(15.000000,4.385650e-01)(20.000000,1.810978e-01)(25.000000,6.144310e-02)(30.000000,1.865700e-02)(35.000000,5.371700e-03)(40.000000,1.589400e-03)(45.000000,5.358000e-04)(50.000000,2.330000e-04)(55.000000,1.516000e-04)(60.000000,1.251000e-04)(65.000000,1.170000e-04)(70.000000,1.143000e-04)(75.000000,1.138000e-04)(80.000000,1.134000e-04)
};
  \addplot[smooth, mark=star,only marks,red] plot coordinates {
(0.000000,9.998746e-01)(5.000000,9.794049e-01)(10.000000,7.745260e-01)(15.000000,3.935506e-01)(20.000000,1.376127e-01)(25.000000,3.938780e-02)(30.000000,1.047220e-02)(35.000000,2.758700e-03)(40.000000,7.519000e-04)(45.000000,2.441000e-04)(50.000000,1.069000e-04)(55.000000,6.620000e-05)(60.000000,5.380000e-05)(65.000000,5.010000e-05)(70.000000,4.880000e-05)(75.000000,4.860000e-05)(80.000000,4.870000e-05)
};

\legend{$W_0=5$cm, $W_0=2$cm, $W_0=1$cm, ,,,High SNR,,,,,,Simulation};

 \draw \boundellipse{ axis cs:67.5,4.3e-02}{10}{1.4};
\node at (axis cs:67.5,3e-001){\scriptsize{$r=2$}};

\draw \boundellipse{ axis cs:67.5,6e-04}{10}{2.8};
\node at (axis cs:67.5,2.3e-005){\scriptsize{$r=1$}};

 \end{axis}
  \end{tikzpicture}
     \caption{OP under SSPA model with ${\rm IBO}=25$ dB for different values of the transmitter beam size $W_0$ with $C_n^2(0)=1\times 10^{-12} {\rm m}^{-\frac{2}{3}}$ under light shadowing conditions for $\xi=1.1$ and $\gamma_{\rm{th}}=0$ dB.}
          \label{fig:OP2}
     \end{center}
  \end{figure}

Fig.~\ref{fig:OP2} illustrates the effect of beam wander associated
with an untracked collimated beam on the outage performance for three different transmitter beam sizes $W_0$ corresponding to 1, 2, and 5 cm. We consider the SSPA model with ${\rm{IBO}}=25$ dB. Based on (\ref{Fried}), a value of $r_0=1.8$ cm is calculated for the Fried's atmospheric coherence width. The effect of the pointing error is fixed at $\xi=1.1$. As clearly seen for this figure, the outage performance under both IM/DD (i.e. $r=2$) and heterodyne (i.e. $r=1$) techniques is reduced when the transmitter beam size increases and it becomes worse when $W_0/r_0>>1$. This is due to the fact that the scintillation index (SI) becomes higher as the ratio $W_0/r_0$ increases as demonstrated in \cite{Andrews}. Indeed, using \cite[p.~517]{Andrews}, the above mentioned transmitter beam sizes correspond to SI$=$0.81, 1.05, and 1.97, respectively for $C_n^2(0)=1\times 10^{-12} {\rm m}^{-\frac{2}{3}}$. Moreover, Fig.~\ref{fig:OP1} indicates that the heterodyne detection always performs better than the IM/DD technique for all SNR range, as expected. Although most of laser SatCom systems are based on the direct detection technique due to its simplicity and ease of deployment \cite{Andrews},
coherent detection for the feeder link is preferred as it offers better spectral efficiency and higher sensitivity, compared to the IM/DD technique \cite{Surof17}. Other outcomes, specifically for the high SNR asymptotic results and the outage floor due to HPA nonlinearity, can be clearly seen similar to Fig.~\ref{fig:OP1} above.

Fig.~\ref{fig:OP3} depicts the effect of changing the nominal ground turbulence levels on the outage performance under both IM/DD and heterodyne techniques without beam wander effect in the case when $\gamma_{\rm{th}}=5$ dB.
  We set the pointing error parameter to $\xi=1.1$ and consider the TWTA model for nonlinear HPA with ${\rm IBO}=25$ dB. We can observe that reducing the ground turbulence level results in better performance under both detection techniques.
This phenomenon is due to the fact that the strength of the optical turbulence decreases as $C_n^2(0)$ becomes smaller.
\begin{figure}[!h]
  \begin{center}
\begin{tikzpicture}[scale=1.2]
    \begin{axis}[font=\footnotesize,
    ymode=log, xlabel= Average Electrical SNR $\mu_r$ (dB), ylabel= Outage Probability,
  xmin=0,xmax=80,ymax=1,ymin=1e-04,
    legend style={nodes=right},legend pos= south west,legend style={nodes={scale=0.65, transform shape}},
      xminorgrids,
    grid style={dotted},
    yminorgrids,
   ]
     \addplot[smooth,blue,mark=triangle*,mark options={solid},every mark/.append style={solid, fill=white}] plot coordinates {
(0.000000,9.971541e-01)(5.000000,9.735451e-01)(10.000000,8.858493e-01)(15.000000,7.122157e-01)(20.000000,4.975632e-01)(25.000000,3.095020e-01)(30.000000,1.800357e-01)(35.000000,1.059470e-01)(40.000000,6.975340e-02)(45.000000,5.476130e-02)(50.000000,4.927290e-02)(55.000000,4.744120e-02)(60.000000,4.683980e-02)(65.000000,4.664460e-02)(70.000000,4.658370e-02)(75.000000,4.656430e-02)(80.000000,4.655840e-02)
};
    \addplot[smooth,green!70!black,mark=*,mark options={solid},every mark/.append style={solid, fill=white}] plot coordinates {
(0.000000,9.993807e-01)(5.000000,9.857865e-01)(10.000000,9.007995e-01)(15.000000,6.950759e-01)(20.000000,4.429716e-01)(25.000000,2.476508e-01)(30.000000,1.321131e-01)(35.000000,7.300120e-02)(40.000000,4.580220e-02)(45.000000,3.460940e-02)(50.000000,3.054620e-02)(55.000000,2.918660e-02)(60.000000,2.874570e-02)(65.000000,2.859890e-02)(70.000000,2.855600e-02)(75.000000,2.854200e-02)(80.000000,2.853690e-02)
};
    \addplot[smooth,cyan,mark=diamond*,mark options={solid},every mark/.append style={solid, fill=white}] plot coordinates {
(0.000000,9.999981e-01)(5.000000,9.984796e-01)(10.000000,9.373603e-01)(15.000000,6.631170e-01)(20.000000,3.584091e-01)(25.000000,1.816273e-01)(30.000000,9.302500e-02)(35.000000,5.014070e-02)(40.000000,3.044940e-02)(45.000000,2.221990e-02)(50.000000,1.918950e-02)(55.000000,1.816650e-02)(60.000000,1.782590e-02)(65.000000,1.771970e-02)(70.000000,1.768110e-02)(75.000000,1.766860e-02)(80.000000,1.766440e-02)
};
     \addplot[smooth,blue,mark=triangle*,mark options={solid},every mark/.append style={solid, fill=white}] plot coordinates {
(0.000000,1.000000e+00)(5.000000,9.999687e-01)(10.000000,9.848470e-01)(15.000000,7.715994e-01)(20.000000,3.596218e-01)(25.000000,1.134555e-01)(30.000000,3.115050e-02)(35.000000,8.894400e-03)(40.000000,3.020600e-03)(45.000000,1.413900e-03)(50.000000,9.513000e-04)(55.000000,8.120000e-04)(60.000000,7.687000e-04)(65.000000,7.557000e-04)(70.000000,7.502000e-04)(75.000000,7.494000e-04)(80.000000,7.489000e-04)
};
    \addplot[smooth,green!70!black,mark=*,mark options={solid},every mark/.append style={solid, fill=white}] plot coordinates {
(0.000000,1.000000e+00)(5.000000,9.999989e-01)(10.000000,9.933526e-01)(15.000000,7.666966e-01)(20.000000,2.981482e-01)(25.000000,8.076870e-02)(30.000000,2.116360e-02)(35.000000,5.998300e-03)(40.000000,2.015300e-03)(45.000000,9.500000e-04)(50.000000,6.385000e-04)(55.000000,5.483000e-04)(60.000000,5.202000e-04)(65.000000,5.112000e-04)(70.000000,5.081000e-04)(75.000000,5.074000e-04)(80.000000,5.072000e-04)
};
    \addplot[smooth,cyan,mark=diamond*,mark options={solid},every mark/.append style={solid, fill=white}] plot coordinates {
(0.000000,1.000000e+00)(5.000000,1.000000e+00)(10.000000,9.996871e-01)(15.000000,7.628131e-01)(20.000000,2.244569e-01)(25.000000,5.733450e-02)(30.000000,1.498850e-02)(35.000000,4.227800e-03)(40.000000,1.425500e-03)(45.000000,6.573000e-04)(50.000000,4.353000e-04)(55.000000,3.730000e-04)(60.000000,3.529000e-04)(65.000000,3.467000e-04)(70.000000,3.445000e-04)(75.000000,3.438000e-04)(80.000000,3.435000e-04)
};
     \addplot[smooth,red,densely dashed,  thick] plot coordinates {
(0.000000,2.287780e+05)(5.000000,1.287600e+04)(10.000000,1.062270e+03)(15.000000,1.057070e+02)(20.000000,1.177330e+01)(25.000000,1.653640e+00)(30.000000,3.872490e-01)(35.000000,1.553490e-01)(40.000000,8.707040e-02)(45.000000,6.221170e-02)(50.000000,5.295560e-02)(55.000000,4.949790e-02)(60.000000,4.816980e-02)(65.000000,4.763690e-02)(70.000000,4.741280e-02)(75.000000,4.731430e-02)(80.000000,4.726940e-02)
};
     \addplot[smooth,red,densely dashed,  thick] plot coordinates {
(0.000000,1.261470e+08)(5.000000,1.949910e+06)(10.000000,5.277020e+04)(15.000000,1.706780e+03)(20.000000,5.852640e+01)(25.000000,2.366090e+00)(30.000000,2.492530e-01)(35.000000,9.435700e-02)(40.000000,5.405050e-02)(45.000000,3.835190e-02)(50.000000,3.229380e-02)(55.000000,3.000050e-02)(60.000000,2.911580e-02)(65.000000,2.876040e-02)(70.000000,2.861090e-02)(75.000000,2.854520e-02)(80.000000,2.851520e-02)
};
     \addplot[smooth,red,densely dashed,  thick] plot coordinates {
(35.000000,6.129840e-02)(40.000000,3.567730e-02)(45.000000,2.473790e-02)(50.000000,2.040800e-02)(55.000000,1.875710e-02)(60.000000,1.812100e-02)(65.000000,1.786650e-02)(70.000000,1.775990e-02)(75.000000,1.771330e-02)(80.000000,1.769210e-02)
};
     \addplot[smooth,red,densely dashed,  thick] plot coordinates {
(50.000000,4.296850e-04)(55.000000,6.449450e-04)(60.000000,7.136570e-04)(65.000000,7.354220e-04)(70.000000,7.422950e-04)(75.000000,7.444650e-04)(80.000000,7.451480e-04)
};
     \addplot[smooth,red,densely dashed,  thick] plot coordinates {
(50.000000,2.448700e-04)(55.000000,4.169960e-04)(60.000000,4.718550e-04)(65.000000,4.892250e-04)(70.000000,4.947130e-04)(75.000000,4.964450e-04)(80.000000,4.969900e-04)
};
     \addplot[smooth,red,densely dashed,  thick] plot coordinates {
(50.000000,1.452840e-04)(55.000000,2.841900e-04)(60.000000,3.284150e-04)(65.000000,3.424160e-04)(70.000000,3.468400e-04)(75.000000,3.482370e-04)(80.000000,3.486780e-04)
};
  \addplot[smooth, mark=star,only marks,red] plot coordinates {
(0.000000,9.971541e-01)(5.000000,9.735451e-01)(10.000000,8.858493e-01)(15.000000,7.122157e-01)(20.000000,4.975632e-01)(25.000000,3.095020e-01)(30.000000,1.800357e-01)(35.000000,1.059470e-01)(40.000000,6.975340e-02)(45.000000,5.476130e-02)(50.000000,4.927290e-02)(55.000000,4.744120e-02)(60.000000,4.683980e-02)(65.000000,4.664460e-02)(70.000000,4.658370e-02)(75.000000,4.656430e-02)(80.000000,4.655840e-02)
};
      \addplot[smooth, mark=star,only marks,red] plot coordinates {
(0.000000,9.993807e-01)(5.000000,9.857865e-01)(10.000000,9.007995e-01)(15.000000,6.950759e-01)(20.000000,4.429716e-01)(25.000000,2.476508e-01)(30.000000,1.321131e-01)(35.000000,7.300120e-02)(40.000000,4.580220e-02)(45.000000,3.460940e-02)(50.000000,3.054620e-02)(55.000000,2.918660e-02)(60.000000,2.874570e-02)(65.000000,2.859890e-02)(70.000000,2.855600e-02)(75.000000,2.854200e-02)(80.000000,2.853690e-02)
};
     \addplot[smooth, mark=star,only marks,red] plot coordinates {
(0.000000,9.999981e-01)(5.000000,9.984796e-01)(10.000000,9.373603e-01)(15.000000,6.631170e-01)(20.000000,3.584091e-01)(25.000000,1.816273e-01)(30.000000,9.302500e-02)(35.000000,5.014070e-02)(40.000000,3.044940e-02)(45.000000,2.221990e-02)(50.000000,1.918950e-02)(55.000000,1.816650e-02)(60.000000,1.782590e-02)(65.000000,1.771970e-02)(70.000000,1.768110e-02)(75.000000,1.766860e-02)(80.000000,1.766440e-02)
};
      \addplot[smooth, mark=star,only marks,red] plot coordinates {
(0.000000,1.000000e+00)(5.000000,9.999687e-01)(10.000000,9.848470e-01)(15.000000,7.715994e-01)(20.000000,3.596218e-01)(25.000000,1.134555e-01)(30.000000,3.115050e-02)(35.000000,8.894400e-03)(40.000000,3.020600e-03)(45.000000,1.413900e-03)(50.000000,9.513000e-04)(55.000000,8.120000e-04)(60.000000,7.687000e-04)(65.000000,7.557000e-04)(70.000000,7.502000e-04)(75.000000,7.494000e-04)(80.000000,7.489000e-04)
};
    \addplot[smooth, mark=star,only marks,red] plot coordinates {
(0.000000,1.000000e+00)(5.000000,9.999989e-01)(10.000000,9.933526e-01)(15.000000,7.666966e-01)(20.000000,2.981482e-01)(25.000000,8.076870e-02)(30.000000,2.116360e-02)(35.000000,5.998300e-03)(40.000000,2.015300e-03)(45.000000,9.500000e-04)(50.000000,6.385000e-04)(55.000000,5.483000e-04)(60.000000,5.202000e-04)(65.000000,5.112000e-04)(70.000000,5.081000e-04)(75.000000,5.074000e-04)(80.000000,5.072000e-04)
};
  \addplot[smooth, mark=star,only marks,red] plot coordinates {
(0.000000,1.000000e+00)(5.000000,1.000000e+00)(10.000000,9.996871e-01)(15.000000,7.628131e-01)(20.000000,2.244569e-01)(25.000000,5.733450e-02)(30.000000,1.498850e-02)(35.000000,4.227800e-03)(40.000000,1.425500e-03)(45.000000,6.573000e-04)(50.000000,4.353000e-04)(55.000000,3.730000e-04)(60.000000,3.529000e-04)(65.000000,3.467000e-04)(70.000000,3.445000e-04)(75.000000,3.438000e-04)(80.000000,3.435000e-04)
};

\legend{$C_n^2(0)=10^{-12} {\rm m}^{-\frac{2}{3}}$, $C_n^2(0)=5 \times 10^{-13} {\rm m}^{-\frac{2}{3}}$, $C_n^2(0)=10^{-13} {\rm m}^{-\frac{2}{3}}$, ,,,High SNR,,,,,,Simulation};

 \draw \boundellipse{ axis cs:72.5,2.854200e-02}{10}{1};
\node at (axis cs:72.5,1e-001){\scriptsize{IM/DD}};

\draw \boundellipse{ axis cs:72.5,5.074000e-04}{10}{0.8};
\node at (axis cs:72.5,1.3e-003){\scriptsize{Heterodyne}};

 \end{axis}
  \end{tikzpicture}
     \caption{OP under TWTA model with ${\rm IBO}=25$ dB for different values of $C_n^2(0)$ without beam wander effect under light shadowing conditions for $\xi=1.1$ and $\gamma_{\rm{th}}=5$ dB with the asymptotic results at high SNR.}
          \label{fig:OP3}
     \end{center}
  \end{figure}
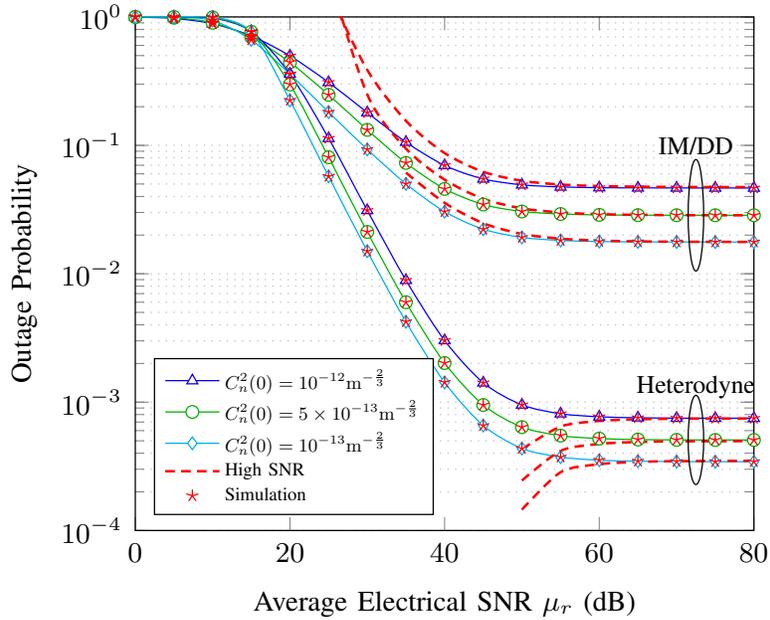
 \begin{figure}[!h]
  \begin{center}
\begin{tikzpicture}[scale=1.2]
    \begin{axis}[font=\footnotesize,
    ymode=log, xlabel= Average Electrical SNR $\mu_r$ (dB), ylabel=  Average Bit-Error Rate,
  xmin=0,xmax=80,ymax=0.5,ymin=1e-05,
    legend style={nodes=right},legend pos= south west,legend style={nodes={scale=1, transform shape}},
      xminorgrids,
    grid style={dotted},
    yminorgrids,
   ]
\addplot[smooth,blue,mark=triangle*,mark options={solid},every mark/.append style={solid, fill=white}] plot coordinates {
(0.000000,4.006397e-01)(5.000000,3.378727e-01)(10.000000,2.638407e-01)(15.000000,2.046677e-01)(20.000000,1.733447e-01)(25.000000,1.609024e-01)(30.000000,1.565991e-01)(35.000000,1.551917e-01)(40.000000,1.547366e-01)(45.000000,1.545912e-01)(50.000000,1.545448e-01)(55.000000,1.545300e-01)(60.000000,1.545255e-01)(65.000000,1.545240e-01)(70.000000,1.545233e-01)(75.000000,1.545233e-01)(80.000000,1.545232e-01)
};

 \addplot[smooth,blue,mark=triangle*,mark options={solid},every mark/.append style={solid, fill=white}] plot coordinates {
(0.000000,3.972157e-01)(5.000000,3.273436e-01)(10.000000,2.333258e-01)(15.000000,1.400466e-01)(20.000000,7.542991e-02)(25.000000,4.090337e-02)(30.000000,2.470358e-02)(35.000000,1.788916e-02)(40.000000,1.533905e-02)(45.000000,1.447471e-02)(50.000000,1.419032e-02)(55.000000,1.409868e-02)(60.000000,1.406987e-02)(65.000000,1.406060e-02)(70.000000,1.405766e-02)(75.000000,1.405680e-02)(80.000000,1.405646e-02)
};

    \addplot[smooth,blue,mark=triangle*,mark options={solid},every mark/.append style={solid, fill=white}] plot coordinates {
(0.000000,3.970311e-01)(5.000000,3.269415e-01)(10.000000,2.324837e-01)(15.000000,1.382820e-01)(20.000000,7.243472e-02)(25.000000,3.635541e-02)(30.000000,1.812712e-02)(35.000000,9.025342e-03)(40.000000,4.490343e-03)(45.000000,2.232696e-03)(50.000000,1.110017e-03)(55.000000,5.516692e-04)(60.000000,2.742854e-04)(65.000000,1.360943e-04)(70.000000,6.766060e-05)(75.000000,3.344951e-05)(80.000000,1.662243e-05)
};

    \addplot[smooth,green!70!black,mark=*,mark options={solid},every mark/.append style={solid, fill=white}] plot coordinates {
(0.000000,3.995425e-01)(5.000000,3.330980e-01)(10.000000,2.492806e-01)(15.000000,1.767141e-01)(20.000000,1.362747e-01)(25.000000,1.199042e-01)(30.000000,1.142116e-01)(35.000000,1.123479e-01)(40.000000,1.117460e-01)(45.000000,1.115548e-01)(50.000000,1.114924e-01)(55.000000,1.114727e-01)(60.000000,1.114669e-01)(65.000000,1.114646e-01)(70.000000,1.114640e-01)(75.000000,1.114637e-01)(80.000000,1.114640e-01)
};

    \addplot[smooth,green!70!black,mark=*,mark options={solid},every mark/.append style={solid, fill=white}] plot coordinates {
(0.000000,3.962906e-01)(5.000000,3.230383e-01)(10.000000,2.184599e-01)(15.000000,1.083523e-01)(20.000000,3.698432e-02)(25.000000,1.000912e-02)(30.000000,2.965709e-03)(35.000000,1.280865e-03)(40.000000,8.322785e-04)(45.000000,7.034243e-04)(50.000000,6.629505e-04)(55.000000,6.513353e-04)(60.000000,6.471629e-04)(65.000000,6.459542e-04)(70.000000,6.455748e-04)(75.000000,6.454328e-04)(80.000000,6.453949e-04)
};
    \addplot[smooth,green!70!black,mark=*,mark options={solid},every mark/.append style={solid, fill=white}] plot coordinates {
(0.000000,3.961135e-01)(5.000000,3.226752e-01)(10.000000,2.176429e-01)(15.000000,1.066462e-01)(20.000000,3.469860e-02)(25.000000,8.068652e-03)(30.000000,1.625704e-03)(35.000000,3.224332e-04)(40.000000,6.451005e-05)(45.000000,1.344224e-05)(50.000000,2.576476e-06)(55.000000,8.255343e-07)(60.000000,1.744691e-07)(65.000000,4.469409e-08)(70.000000,1.512771e-09)(75.000000,8.098553e-12)(80.000000,4.405263e-22)
};


\addplot[smooth, mark=star,only marks,red] plot coordinates {
(0.000000,4.006397e-01)(5.000000,3.378727e-01)(10.000000,2.638407e-01)(15.000000,2.046677e-01)(20.000000,1.733447e-01)(25.000000,1.609024e-01)(30.000000,1.565991e-01)(35.000000,1.551917e-01)(40.000000,1.547366e-01)(45.000000,1.545912e-01)(50.000000,1.545448e-01)(55.000000,1.545300e-01)(60.000000,1.545255e-01)(65.000000,1.545240e-01)(70.000000,1.545233e-01)(75.000000,1.545233e-01)(80.000000,1.545232e-01)
};

\addplot[smooth, mark=star,only marks,red] plot coordinates {
(0.000000,3.972157e-01)(5.000000,3.273436e-01)(10.000000,2.333258e-01)(15.000000,1.400466e-01)(20.000000,7.542991e-02)(25.000000,4.090337e-02)(30.000000,2.470358e-02)(35.000000,1.788916e-02)(40.000000,1.533905e-02)(45.000000,1.447471e-02)(50.000000,1.419032e-02)(55.000000,1.409868e-02)(60.000000,1.406987e-02)(65.000000,1.406060e-02)(70.000000,1.405766e-02)(75.000000,1.405680e-02)(80.000000,1.405646e-02)
};

\addplot[smooth, mark=star,only marks,red] plot coordinates {
(0.000000,3.970311e-01)(5.000000,3.269415e-01)(10.000000,2.324837e-01)(15.000000,1.382820e-01)(20.000000,7.243472e-02)(25.000000,3.635541e-02)(30.000000,1.812712e-02)(35.000000,9.025342e-03)(40.000000,4.490343e-03)(45.000000,2.232696e-03)(50.000000,1.110017e-03)(55.000000,5.516692e-04)(60.000000,2.742854e-04)(65.000000,1.360943e-04)(70.000000,6.766060e-05)(75.000000,3.344951e-05)(80.000000,1.662243e-05)
};

\addplot[smooth, mark=star,only marks,red] plot coordinates {
(0.000000,3.995425e-01)(5.000000,3.330980e-01)(10.000000,2.492806e-01)(15.000000,1.767141e-01)(20.000000,1.362747e-01)(25.000000,1.199042e-01)(30.000000,1.142116e-01)(35.000000,1.123479e-01)(40.000000,1.117460e-01)(45.000000,1.115548e-01)(50.000000,1.114924e-01)(55.000000,1.114727e-01)(60.000000,1.114669e-01)(65.000000,1.114646e-01)(70.000000,1.114640e-01)(75.000000,1.114637e-01)(80.000000,1.114640e-01)
};

\addplot[smooth, mark=star,only marks,red] plot coordinates {
(0.000000,3.962906e-01)(5.000000,3.230383e-01)(10.000000,2.184599e-01)(15.000000,1.083523e-01)(20.000000,3.698432e-02)(25.000000,1.000912e-02)(30.000000,2.965709e-03)(35.000000,1.280865e-03)(40.000000,8.322785e-04)(45.000000,7.034243e-04)(50.000000,6.629505e-04)(55.000000,6.513353e-04)(60.000000,6.471629e-04)(65.000000,6.459542e-04)(70.000000,6.455748e-04)(75.000000,6.454328e-04)(80.000000,6.453949e-04)
};
\addplot[smooth, mark=star,only marks,red] plot coordinates {
(0.000000,3.961135e-01)(5.000000,3.226752e-01)(10.000000,2.176429e-01)(15.000000,1.066462e-01)(20.000000,3.469860e-02)(25.000000,8.068652e-03)(30.000000,1.625704e-03)(35.000000,3.224332e-04)(40.000000,6.451005e-05)(45.000000,1.344224e-05)(50.000000,2.576476e-06)(55.000000,8.255343e-07)(60.000000,1.744691e-07)(65.000000,4.469409e-08)(70.000000,1.512771e-09)(75.000000,8.098553e-12)(80.000000,4.405263e-22)
};

\legend{$\xi=1.1$,,, $\xi=1.7$,,,Simulation};

 \draw \boundellipse{ axis cs:67.5,1.3e-01}{10}{0.4};
\node at (axis cs:67.5,2.8e-01){\scriptsize{IBO$=$10dB}};

\draw \boundellipse{ axis cs:67.5,3e-03}{10}{2};
\node at (axis cs:67.5,3e-02){\scriptsize{IBO$=$20dB}};

\draw \boundellipse{ axis cs:49,2e-04}{150}{0.2};
\node at (axis cs:72,2.1e-04){\scriptsize{Linear PA}};

 \end{axis}
  \end{tikzpicture}
     \caption{Average BER with OOK under IM/DD for different TWTA IBOs and pointing errors with $C_n^2(0)=1\times 10^{-13} {\rm m}^{-\frac{2}{3}}$ under light shadowing conditions.}
          \label{fig:BER1}
     \end{center}
  \end{figure}
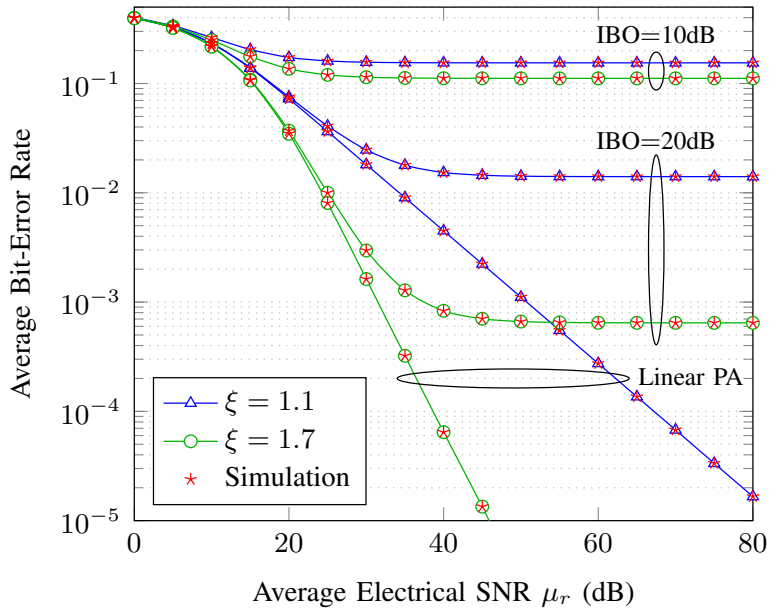

  The average BER performance with OOK modulation under TWTA model with 10 and 20 dB IBOs and different values of the pointing error parameter $\xi$ without beam wander effect is shown in Fig.~\ref{fig:BER1}.
  Under the same conditions, results of the linear PA are also plotted in Fig.~\ref{fig:BER1}.
Similar to the outage probability analysis, the same floor effect is observed here and as clearly seen, its level becomes higher as the value of IBO gets smaller and it vanishes when linear PA is employed. Moreover, the average BER performance improves as IBO increases for both values of $\xi$. This can be attributed to the fact that larger IBO values are associated with higher input power saturation levels $A_{\rm{sat}}$ and consequently result in lower nonlinear distortion caused by HPA. Fig.~\ref{fig:BER1} also illustrates the effect of the pointing error on the BER performance. We can observe that for higher values of $\xi$, the effect of the pointing error is less severe and the average BER gets better, especially for higher values of IBO.

To further illustrate the effect of IBO, the average BER with OOK modulation under different TWTA and SSPA IBO values is depicted in Fig.~\ref{fig:BER2} for $\xi=1.1$. As can be seen, the distortion caused by the amplifier's nonlinearity under both TWTA and SSPA models results in a degradation of the average BER performance, which becomes larger for lower values of IBO and a BER floor is introduced under both TWTA and SSPA especially at high SNR. As can be seen, the distortion caused by the amplifier's nonlinearity under both TWTA and SSPA models results in a degradation of the average BER performance, which becomes larger for lower values of IBO and a BER floor is introduced under both TWTA and SSPA especially at high SNR. Indeed, the effect of IBO on the average BER performance becomes more significant with the increase of the average electrical SNR. At low values of $\mu_r$, the IBO has a negligible impact on the performance and the system operates efficiently. This can be clearly observed from this figure as the average BER is almost the same for all values of IBO, especially for an average electrical SNR less than 15 dB. As $\mu_r$ increases, the IBO parameter becomes more involved and the average BER performance improves as IBO gets larger.
 \begin{figure}[!h]
  \begin{center}
\begin{tikzpicture}[scale=1.2]
    \begin{axis}[font=\footnotesize,
    ymode=log, xlabel= Average Electrical SNR $\mu_r$ (dB), ylabel=  Average Bit-Error Rate,
  xmin=0,xmax=80,ymax=0.5,ymin=1e-04,
    legend style={nodes=right},legend pos= south west,legend style={nodes={scale=1, transform shape}},
      xminorgrids,
    grid style={dotted},
    yminorgrids,
   ]

\addplot[smooth,blue,mark=triangle*,mark options={solid},every mark/.append style={solid, fill=white}] plot coordinates {
(0.000000,3.465466e-01)(5.000000,2.859622e-01)(10.000000,2.321740e-01)(15.000000,1.967464e-01)(20.000000,1.794983e-01)(25.000000,1.727389e-01)(30.000000,1.703802e-01)(35.000000,1.695876e-01)(40.000000,1.693282e-01)(45.000000,1.692441e-01)(50.000000,1.692157e-01)(55.000000,1.692072e-01)(60.000000,1.692044e-01)(65.000000,1.692032e-01)(70.000000,1.692030e-01)(75.000000,1.692030e-01)(80.000000,1.692029e-01)
};

 \addplot[smooth,green!70!black,mark=*,mark options={solid},every mark/.append style={solid, fill=white}] plot coordinates {
(0.000000,3.413735e-01)(5.000000,2.754788e-01)(10.000000,2.108544e-01)(15.000000,1.596736e-01)(20.000000,1.283212e-01)(25.000000,1.135181e-01)(30.000000,1.077817e-01)(35.000000,1.057858e-01)(40.000000,1.051240e-01)(45.000000,1.049107e-01)(50.000000,1.048395e-01)(55.000000,1.048166e-01)(60.000000,1.048094e-01)(65.000000,1.048070e-01)(70.000000,1.048062e-01)(75.000000,1.048060e-01)(80.000000,1.048059e-01)
};

\addplot[smooth,blue,mark=triangle*,mark options={solid},every mark/.append style={solid, fill=white}] plot coordinates {
(0.000000,3.385746e-01)(5.000000,2.705243e-01)(10.000000,2.009171e-01)(15.000000,1.395032e-01)(20.000000,9.253601e-02)(25.000000,6.137666e-02)(30.000000,4.391998e-02)(35.000000,3.583518e-02)(40.000000,3.269381e-02)(45.000000,3.160133e-02)(50.000000,3.123495e-02)(55.000000,3.111701e-02)(60.000000,3.107730e-02)(65.000000,3.106497e-02)(70.000000,3.106109e-02)(75.000000,3.105989e-02)(80.000000,3.105937e-02)
};

 \addplot[smooth,green!70!black,mark=*,mark options={solid},every mark/.append style={solid, fill=white}] plot coordinates {
(0.000000,3.381893e-01)(5.000000,2.698963e-01)(10.000000,1.999070e-01)(15.000000,1.378887e-01)(20.000000,8.967819e-02)(25.000000,5.618198e-02)(30.000000,3.532371e-02)(35.000000,2.395501e-02)(40.000000,1.869133e-02)(45.000000,1.662828e-02)(50.000000,1.590514e-02)(55.000000,1.566469e-02)(60.000000,1.558395e-02)(65.000000,1.555832e-02)(70.000000,1.554928e-02)(75.000000,1.554674e-02)(80.000000,1.554592e-02)
};


    \addplot[smooth,blue,mark=triangle*,mark options={solid},every mark/.append style={solid, fill=white}] plot coordinates {
(0.000000,3.381272e-01)(5.000000,2.698200e-01)(10.000000,1.997558e-01)(15.000000,1.375854e-01)(20.000000,8.905719e-02)(25.000000,5.496000e-02)(30.000000,3.320810e-02)(35.000000,2.061040e-02)(40.000000,1.419479e-02)(45.000000,1.140568e-02)(50.000000,1.035389e-02)(55.000000,9.991602e-03)(60.000000,9.868489e-03)(65.000000,9.830245e-03)(70.000000,9.818004e-03)(75.000000,9.814058e-03)(80.000000,9.812568e-03)
};

    \addplot[smooth,green!70!black,mark=*,mark options={solid},every mark/.append style={solid, fill=white}] plot coordinates {
(0.000000,3.380752e-01)(5.000000,2.698476e-01)(10.000000,1.997572e-01)(15.000000,1.376128e-01)(20.000000,8.887978e-02)(25.000000,5.450387e-02)(30.000000,3.222931e-02)(35.000000,1.877284e-02)(40.000000,1.118341e-02)(45.000000,7.362716e-03)(50.000000,5.686788e-03)(55.000000,5.040172e-03)(60.000000,4.816185e-03)(65.000000,4.742883e-03)(70.000000,4.717909e-03)(75.000000,4.709658e-03)(80.000000,4.707157e-03)
};

    \addplot[smooth,cyan,mark=diamond*,mark options={solid},every mark/.append style={solid, fill=white}] plot coordinates {
(0.000000,3.378808e-01)(5.000000,2.695446e-01)(10.000000,1.994336e-01)(15.000000,1.372015e-01)(20.000000,8.849136e-02)(25.000000,5.404957e-02)(30.000000,3.163385e-02)(35.000000,1.789876e-02)(40.000000,9.847427e-03)(45.000000,5.290841e-03)(50.000000,2.805926e-03)(55.000000,1.477438e-03)(60.000000,7.719534e-04)(65.000000,4.002531e-04)(70.000000,2.107625e-04)(75.000000,1.078271e-04)(80.000000,5.411274e-05)
};
    \addplot[smooth,red,densely dashed,  thick] plot coordinates {
(25.000000,1.092190e-02)(30.000000,3.445110e-02)(35.000000,3.569350e-02)(40.000000,3.362610e-02)(45.000000,3.189050e-02)(50.000000,3.081300e-02)(55.000000,3.020370e-02)(60.000000,2.987370e-02)(65.000000,2.969930e-02)(70.000000,2.960850e-02)(75.000000,2.956180e-02)(80.000000,2.953780e-02)
};

    \addplot[smooth,red,densely dashed,  thick] plot coordinates {
(25.000000,9.494010e-03)(30.000000,2.800980e-02)(35.000000,2.536990e-02)(40.000000,2.099260e-02)(45.000000,1.823490e-02)(50.000000,1.678390e-02)(55.000000,1.605040e-02)(60.000000,1.568100e-02)(65.000000,1.549450e-02)(70.000000,1.540030e-02)(75.000000,1.535260e-02)(80.000000,1.532840e-02)
};


    \addplot[smooth,red,densely dashed,  thick] plot coordinates {
(25.000000,9.252960e-03)(30.000000,2.639420e-02)(35.000000,2.234980e-02)(40.000000,1.673200e-02)(45.000000,1.321340e-02)(50.000000,1.142620e-02)(55.000000,1.057000e-02)(60.000000,1.015950e-02)(65.000000,9.959850e-03)(70.000000,9.861390e-03)(75.000000,9.812420e-03)(80.000000,9.787930e-03)
};

    \addplot[smooth,red,densely dashed,  thick] plot coordinates {
(25.000000,9.221740e-03)(30.000000,2.547600e-02)(35.000000,2.045260e-02)(40.000000,1.369250e-02)(45.000000,9.153550e-03)(50.000000,6.741730e-03)(55.000000,5.607970e-03)(60.000000,5.095460e-03)(65.000000,4.861350e-03)(70.000000,4.751730e-03)(75.000000,4.699160e-03)(80.000000,4.673570e-03)
};
    \addplot[smooth,red,densely dashed,  thick] plot coordinates {
(30.000000,1.078960e-02)(35.000000,1.096430e-02)(40.000000,7.435540e-03)(45.000000,4.412500e-03)(50.000000,2.457360e-03)(55.000000,1.320870e-03)(60.000000,6.947510e-04)(65.000000,3.602970e-04)(70.000000,1.850540e-04)(75.000000,9.439950e-05)(80.000000,4.791510e-05)
};


\addplot[smooth, mark=star,only marks,red] plot coordinates {(0.000000,3.465466e-01)(5.000000,2.859622e-01)(10.000000,2.321740e-01)(15.000000,1.967464e-01)(20.000000,1.794983e-01)(25.000000,1.727389e-01)(30.000000,1.703802e-01)(35.000000,1.695876e-01)(40.000000,1.693282e-01)(45.000000,1.692441e-01)(50.000000,1.692157e-01)(55.000000,1.692072e-01)(60.000000,1.692044e-01)(65.000000,1.692032e-01)(70.000000,1.692030e-01)(75.000000,1.692030e-01)(80.000000,1.692029e-01)
};

\addplot[smooth, mark=star,only marks,red] plot coordinates {(0.000000,3.413735e-01)(5.000000,2.754788e-01)(10.000000,2.108544e-01)(15.000000,1.596736e-01)(20.000000,1.283212e-01)(25.000000,1.135181e-01)(30.000000,1.077817e-01)(35.000000,1.057858e-01)(40.000000,1.051240e-01)(45.000000,1.049107e-01)(50.000000,1.048395e-01)(55.000000,1.048166e-01)(60.000000,1.048094e-01)(65.000000,1.048070e-01)(70.000000,1.048062e-01)(75.000000,1.048060e-01)(80.000000,1.048059e-01)
};
\addplot[smooth, mark=star,only marks,red] plot coordinates {
(0.000000,3.385746e-01)(5.000000,2.705243e-01)(10.000000,2.009171e-01)(15.000000,1.395032e-01)(20.000000,9.253601e-02)(25.000000,6.137666e-02)(30.000000,4.391998e-02)(35.000000,3.583518e-02)(40.000000,3.269381e-02)(45.000000,3.160133e-02)(50.000000,3.123495e-02)(55.000000,3.111701e-02)(60.000000,3.107730e-02)(65.000000,3.106497e-02)(70.000000,3.106109e-02)(75.000000,3.105989e-02)(80.000000,3.105937e-02)
};

\addplot[smooth, mark=star,only marks,red] plot coordinates {
(0.000000,3.381893e-01)(5.000000,2.698963e-01)(10.000000,1.999070e-01)(15.000000,1.378887e-01)(20.000000,8.967819e-02)(25.000000,5.618198e-02)(30.000000,3.532371e-02)(35.000000,2.395501e-02)(40.000000,1.869133e-02)(45.000000,1.662828e-02)(50.000000,1.590514e-02)(55.000000,1.566469e-02)(60.000000,1.558395e-02)(65.000000,1.555832e-02)(70.000000,1.554928e-02)(75.000000,1.554674e-02)(80.000000,1.554592e-02)
};


\addplot[smooth, mark=star,only marks,red] plot coordinates {
(0.000000,3.381272e-01)(5.000000,2.698200e-01)(10.000000,1.997558e-01)(15.000000,1.375854e-01)(20.000000,8.905719e-02)(25.000000,5.496000e-02)(30.000000,3.320810e-02)(35.000000,2.061040e-02)(40.000000,1.419479e-02)(45.000000,1.140568e-02)(50.000000,1.035389e-02)(55.000000,9.991602e-03)(60.000000,9.868489e-03)(65.000000,9.830245e-03)(70.000000,9.818004e-03)(75.000000,9.814058e-03)(80.000000,9.812568e-03)
};

\addplot[smooth, mark=star,only marks,red] plot coordinates {
(0.000000,3.380752e-01)(5.000000,2.698476e-01)(10.000000,1.997572e-01)(15.000000,1.376128e-01)(20.000000,8.887978e-02)(25.000000,5.450387e-02)(30.000000,3.222931e-02)(35.000000,1.877284e-02)(40.000000,1.118341e-02)(45.000000,7.362716e-03)(50.000000,5.686788e-03)(55.000000,5.040172e-03)(60.000000,4.816185e-03)(65.000000,4.742883e-03)(70.000000,4.717909e-03)(75.000000,4.709658e-03)(80.000000,4.707157e-03)
};
\addplot[smooth, mark=star,only marks,red] plot coordinates {
(0.000000,3.378808e-01)(5.000000,2.695446e-01)(10.000000,1.994336e-01)(15.000000,1.372015e-01)(20.000000,8.849136e-02)(25.000000,5.404957e-02)(30.000000,3.163385e-02)(35.000000,1.789876e-02)(40.000000,9.847427e-03)(45.000000,5.290841e-03)(50.000000,2.805926e-03)(55.000000,1.477438e-03)(60.000000,7.719534e-04)(65.000000,4.002531e-04)(70.000000,2.107625e-04)(75.000000,1.078271e-04)(80.000000,5.411274e-05)
};

\legend{TWTA,SSPA,,,,,Linear PA,,,High SNR,,,,,Simulation};
 \draw \boundellipse{ axis cs:67.5,1.3e-01}{10}{0.5};
\node at (axis cs:67.5,2.8e-01){\scriptsize{IBO$=$10dB}};

\draw \boundellipse{ axis cs:67.5,2.2e-02}{10}{0.55};
\node at (axis cs:67.5,5.2e-02){\scriptsize{IBO$=$20dB}};

\draw \boundellipse{ axis cs:67.5,6.8e-03}{10}{0.55};
\node at (axis cs:67.5,3e-03){\scriptsize{IBO$=$25dB}};

 \end{axis}
  \end{tikzpicture}
     \caption{Average BER with OOK under IM/DD under TWTA and SSPA models with different values of IBO for $\xi=1.1$
     and $C_n^2(0)=1\times 10^{-12} {\rm m}^{-\frac{2}{3}}$ with beam wander effect under light shadowing conditions.}
          \label{fig:BER2}
     \end{center}
  \end{figure}
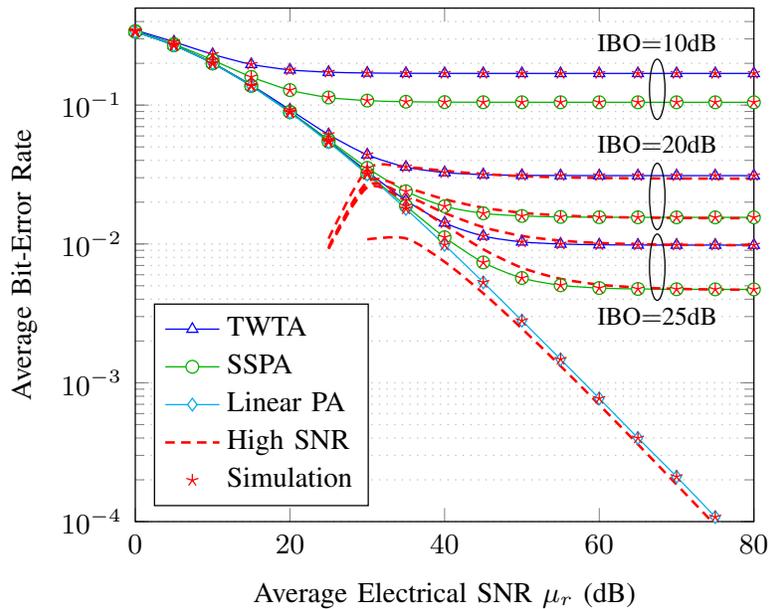\noindent
Moreover, for all values of IBO, SSPA performs better than TWTA but its performance is still inferior to that of the linear PA. Additionally, it can be concluded that, it is necessary to use large values of IBO to obtain similar performance to the linear PA at least up to 40 dB of average electrical SNR.
Furthermore, it can be observed that in the high SNR regime, the asymptotic expression of the average BER derived in (\ref{asymBER}) converges perfectly to the exact result proving the tightness of this asymptotic result.

\begin{figure}[!h]
  \begin{center}
\begin{tikzpicture}[scale=1.2]
    \begin{axis}[xtick=data,font=\footnotesize,
    ymode=log, xlabel= Average Electrical SNR $\mu_r$ (dB), ylabel= Average Bit-Error Rate,
  xmin=0,xmax=80,ymin=1e-04, ymax=1,
    legend style={nodes=right},legend pos= south west,legend style={font=\footnotesize},
    xtick=data,
      xminorgrids,
    grid style={dotted},
    yminorgrids,
   ]
     \addplot[smooth,blue,mark=triangle*,mark options={solid},every mark/.append style={solid, fill=white}] plot coordinates {
(0.000000,1.030925e+00)(5.000000,9.438147e-01)(10.000000,8.276159e-01)(15.000000,6.952013e-01)(20.000000,5.623996e-01)(25.000000,4.412736e-01)(30.000000,3.399540e-01)(35.000000,2.627730e-01)(40.000000,2.102478e-01)(45.000000,1.802234e-01)(50.000000,1.665421e-01)(55.000000,1.613774e-01)(60.000000,1.596347e-01)(65.000000,1.590632e-01)(70.000000,1.588802e-01)(75.000000,1.588225e-01)(80.000000,1.588035e-01)
};
    \addplot[smooth,green!70!black,mark=*,mark options={solid},every mark/.append style={solid, fill=white}] plot coordinates {
(0.000000,8.785730e-01)(5.000000,7.998479e-01)(10.000000,6.947107e-01)(15.000000,5.760539e-01)(20.000000,4.596853e-01)(25.000000,3.565339e-01)(30.000000,2.728520e-01)(35.000000,2.104137e-01)(40.000000,1.683108e-01)(45.000000,1.442722e-01)(50.000000,1.333272e-01)(55.000000,1.292056e-01)(60.000000,1.278070e-01)(65.000000,1.273508e-01)(70.000000,1.272061e-01)(75.000000,1.271604e-01)(80.000000,1.271452e-01)
};
    \addplot[smooth,brown,mark=diamond*,mark options={solid},every mark/.append style={solid, fill=white}] plot coordinates {
(0.000000,6.603444e-01)(5.000000,6.024632e-01)(10.000000,5.247118e-01)(15.000000,4.347936e-01)(20.000000,3.445379e-01)(25.000000,2.650646e-01)(30.000000,2.018950e-01)(35.000000,1.553762e-01)(40.000000,1.241472e-01)(45.000000,1.063621e-01)(50.000000,9.827686e-02)(55.000000,9.523226e-02)(60.000000,9.419972e-02)(65.000000,9.386443e-02)(70.000000,9.375632e-02)(75.000000,9.372194e-02)(80.000000,9.371100e-02)
};
    \addplot[smooth,cyan,mark=pentagon*,mark options={solid},every mark/.append style={solid, fill=white}] plot coordinates {
(0.000000,4.072552e-01)(5.000000,3.518626e-01)(10.000000,2.858070e-01)(15.000000,2.217275e-01)(20.000000,1.682544e-01)(25.000000,1.268467e-01)(30.000000,9.590092e-02)(35.000000,7.360592e-02)(40.000000,5.877919e-02)(45.000000,5.035706e-02)(50.000000,4.653715e-02)(55.000000,4.509751e-02)(60.000000,4.460750e-02)(65.000000,4.444872e-02)(70.000000,4.439791e-02)(75.000000,4.438165e-02)(80.000000,4.437650e-02)
};

     \addplot[smooth,blue,mark=triangle*,mark options={solid},every mark/.append style={solid, fill=white}] plot coordinates {
(0.000000,9.886308e-01)(5.000000,8.665937e-01)(10.000000,6.955518e-01)(15.000000,5.008351e-01)(20.000000,3.232962e-01)(25.000000,1.865378e-01)(30.000000,9.398660e-02)(35.000000,4.197689e-02)(40.000000,1.886065e-02)(45.000000,1.032587e-02)(50.000000,7.480837e-03)(55.000000,6.559395e-03)(60.000000,6.267064e-03)(65.000000,6.174001e-03)(70.000000,6.143740e-03)(75.000000,6.134301e-03)(80.000000,6.131366e-03)
};
    \addplot[smooth,green!70!black,mark=*,mark options={solid},every mark/.append style={solid, fill=white}] plot coordinates {
(0.000000,8.411772e-01)(5.000000,7.313160e-01)(10.000000,5.761380e-01)(15.000000,4.002921e-01)(20.000000,2.446309e-01)(25.000000,1.309194e-01)(30.000000,6.054385e-02)(35.000000,2.510393e-02)(40.000000,1.071487e-02)(45.000000,5.714718e-03)(50.000000,4.090778e-03)(55.000000,3.570872e-03)(60.000000,3.406410e-03)(65.000000,3.353400e-03)(70.000000,3.337156e-03)(75.000000,3.331976e-03)(80.000000,3.330150e-03)
};
    \addplot[smooth,brown,mark=diamond*,mark options={solid},every mark/.append style={solid, fill=white}] plot coordinates {
(0.000000,6.325392e-01)(5.000000,5.516609e-01)(10.000000,4.375476e-01)(15.000000,3.057060e-01)(20.000000,1.823143e-01)(25.000000,8.984740e-02)(30.000000,3.700040e-02)(35.000000,1.386561e-02)(40.000000,5.537633e-03)(45.000000,2.849935e-03)(50.000000,2.005045e-03)(55.000000,1.742446e-03)(60.000000,1.659811e-03)(65.000000,1.633560e-03)(70.000000,1.625124e-03)(75.000000,1.622467e-03)(80.000000,1.621693e-03)
};
    \addplot[smooth,cyan,mark=pentagon*,mark options={solid},every mark/.append style={solid, fill=white}] plot coordinates {
(0.000000,3.776739e-01)(5.000000,2.978043e-01)(10.000000,1.964127e-01)(15.000000,1.026590e-01)(20.000000,4.274518e-02)(25.000000,1.494992e-02)(30.000000,4.752204e-03)(35.000000,1.500810e-03)(40.000000,5.424761e-04)(45.000000,2.643387e-04)(50.000000,1.813753e-04)(55.000000,1.554188e-04)(60.000000,1.475505e-04)(65.000000,1.450872e-04)(70.000000,1.440788e-04)(75.000000,1.438388e-04)(80.000000,1.437625e-04)
};
     \addplot[smooth,red,densely dashed, thick] plot coordinates {
(25.000000,4.827380e-01)(30.000000,5.075900e-01)(35.000000,3.905710e-01)(40.000000,3.056160e-01)(45.000000,2.515500e-01)(50.000000,2.199680e-01)(55.000000,2.014400e-01)(60.000000,1.896770e-01)(65.000000,1.816020e-01)(70.000000,1.757930e-01)(75.000000,1.715150e-01)(80.000000,1.683320e-01)
};
    \addplot[smooth,red,densely dashed, thick] plot coordinates {
(0.000000,-3.625360e+06)(5.000000,-1.208130e+05)(10.000000,-2.783770e+03)(15.000000,-6.272770e+01)(20.000000,-7.655530e-01)(25.000000,5.073150e-01)(30.000000,4.097410e-01)(35.000000,3.126940e-01)(40.000000,2.446130e-01)(45.000000,2.013620e-01)(50.000000,1.760940e-01)(55.000000,1.612670e-01)(60.000000,1.518520e-01)(65.000000,1.453890e-01)(70.000000,1.407390e-01)(75.000000,1.373140e-01)(80.000000,1.347670e-01)
};
   \addplot[smooth,red,densely dashed,  thick] plot coordinates {
(0.000000,-3.302750e+05)(5.000000,-7.664260e+03)(10.000000,-1.738830e+02)(15.000000,-3.331120e+00)(20.000000,4.357110e-01)(25.000000,3.992060e-01)(30.000000,3.025670e-01)(35.000000,2.303050e-01)(40.000000,1.801580e-01)(45.000000,1.483170e-01)(50.000000,1.297120e-01)(55.000000,1.187940e-01)(60.000000,1.118590e-01)(65.000000,1.070990e-01)(70.000000,1.036740e-01)(75.000000,1.011510e-01)(80.000000,9.927460e-02)
};
 \addplot[smooth,red,densely dashed, thick] plot coordinates {
(0.000000,-2.317790e+02)(5.000000,-4.760190e+00)(10.000000,3.215470e-01)(15.000000,3.359090e-01)(20.000000,2.542240e-01)(25.000000,1.906950e-01)(30.000000,1.436060e-01)(35.000000,1.093650e-01)(40.000000,8.559850e-02)(45.000000,7.049060e-02)(50.000000,6.165630e-02)(55.000000,5.646930e-02)(60.000000,5.317410e-02)(65.000000,5.091150e-02)(70.000000,4.928340e-02)(75.000000,4.808450e-02)(80.000000,4.719240e-02)
};

     \addplot[smooth,red,densely dashed, thick] plot coordinates {
(45.000000,1.970840e-03)(50.000000,4.778190e-03)(55.000000,5.617050e-03)(60.000000,5.877580e-03)(65.000000,5.959350e-03)(70.000000,5.985090e-03)(75.000000,5.993180e-03)(80.000000,5.995730e-03)
};
    \addplot[smooth,red,densely dashed, thick] plot coordinates {
(45.000000,1.187360e-03)(50.000000,2.627160e-03)(55.000000,3.071960e-03)(60.000000,3.211560e-03)(65.000000,3.255510e-03)(70.000000,3.269350e-03)(75.000000,3.273700e-03)(80.000000,3.275070e-03)
};
   \addplot[smooth,red,densely dashed,  thick] plot coordinates {
(45.000000,5.571080e-04)(50.000000,1.277930e-03)(55.000000,1.507590e-03)(60.000000,1.580370e-03)(65.000000,1.603350e-03)(70.000000,1.610590e-03)(75.000000,1.612870e-03)(80.000000,1.613590e-03)
};
 \addplot[smooth,red,densely dashed, thick] plot coordinates {
(0.000000,-4.502910e+04)(5.000000,-1.031220e+03)(10.000000,-2.477770e+01)(15.000000,-7.937540e-01)(20.000000,-6.411240e-02)(25.000000,-1.331030e-02)(30.000000,-3.953090e-03)(35.000000,-1.237120e-03)(40.000000,-3.228190e-04)(45.000000,-9.579600e-06)(50.000000,9.473890e-05)(55.000000,1.284920e-04)(60.000000,1.392430e-04)(65.000000,1.426450e-04)(70.000000,1.437190e-04)(75.000000,1.440580e-04)(80.000000,1.441650e-04)
};
\addplot[smooth, mark=star,only marks,red] plot coordinates {
(0.000000,1.030925e+00)(5.000000,9.438147e-01)(10.000000,8.276159e-01)(15.000000,6.952013e-01)(20.000000,5.623996e-01)(25.000000,4.412736e-01)(30.000000,3.399540e-01)(35.000000,2.627730e-01)(40.000000,2.102478e-01)(45.000000,1.802234e-01)(50.000000,1.665421e-01)(55.000000,1.613774e-01)(60.000000,1.596347e-01)(65.000000,1.590632e-01)(70.000000,1.588802e-01)(75.000000,1.588225e-01)(80.000000,1.588035e-01)
};
   \addplot[smooth, mark=star,only marks,red] plot coordinates {
(0.000000,8.785730e-01)(5.000000,7.998479e-01)(10.000000,6.947107e-01)(15.000000,5.760539e-01)(20.000000,4.596853e-01)(25.000000,3.565339e-01)(30.000000,2.728520e-01)(35.000000,2.104137e-01)(40.000000,1.683108e-01)(45.000000,1.442722e-01)(50.000000,1.333272e-01)(55.000000,1.292056e-01)(60.000000,1.278070e-01)(65.000000,1.273508e-01)(70.000000,1.272061e-01)(75.000000,1.271604e-01)(80.000000,1.271452e-01)
};
   \addplot[smooth, mark=star,only marks,red] plot coordinates {
(0.000000,6.603444e-01)(5.000000,6.024632e-01)(10.000000,5.247118e-01)(15.000000,4.347936e-01)(20.000000,3.445379e-01)(25.000000,2.650646e-01)(30.000000,2.018950e-01)(35.000000,1.553762e-01)(40.000000,1.241472e-01)(45.000000,1.063621e-01)(50.000000,9.827686e-02)(55.000000,9.523226e-02)(60.000000,9.419972e-02)(65.000000,9.386443e-02)(70.000000,9.375632e-02)(75.000000,9.372194e-02)(80.000000,9.371100e-02)
};
   \addplot[smooth, mark=star,only marks,red] plot coordinates {
(0.000000,4.072552e-01)(5.000000,3.518626e-01)(10.000000,2.858070e-01)(15.000000,2.217275e-01)(20.000000,1.682544e-01)(25.000000,1.268467e-01)(30.000000,9.590092e-02)(35.000000,7.360592e-02)(40.000000,5.877919e-02)(45.000000,5.035706e-02)(50.000000,4.653715e-02)(55.000000,4.509751e-02)(60.000000,4.460750e-02)(65.000000,4.444872e-02)(70.000000,4.439791e-02)(75.000000,4.438165e-02)(80.000000,4.437650e-02)
};
\addplot[smooth, mark=star,only marks,red] plot coordinates {
(0.000000,9.886308e-01)(5.000000,8.665937e-01)(10.000000,6.955518e-01)(15.000000,5.008351e-01)(20.000000,3.232962e-01)(25.000000,1.865378e-01)(30.000000,9.398660e-02)(35.000000,4.197689e-02)(40.000000,1.886065e-02)(45.000000,1.032587e-02)(50.000000,7.480837e-03)(55.000000,6.559395e-03)(60.000000,6.267064e-03)(65.000000,6.174001e-03)(70.000000,6.143740e-03)(75.000000,6.134301e-03)(80.000000,6.131366e-03)
};
   \addplot[smooth, mark=star,only marks,red] plot coordinates {
(0.000000,8.411772e-01)(5.000000,7.313160e-01)(10.000000,5.761380e-01)(15.000000,4.002921e-01)(20.000000,2.446309e-01)(25.000000,1.309194e-01)(30.000000,6.054385e-02)(35.000000,2.510393e-02)(40.000000,1.071487e-02)(45.000000,5.714718e-03)(50.000000,4.090778e-03)(55.000000,3.570872e-03)(60.000000,3.406410e-03)(65.000000,3.353400e-03)(70.000000,3.337156e-03)(75.000000,3.331976e-03)(80.000000,3.330150e-03)
};
   \addplot[smooth, mark=star,only marks,red] plot coordinates {
(0.000000,6.325392e-01)(5.000000,5.516609e-01)(10.000000,4.375476e-01)(15.000000,3.057060e-01)(20.000000,1.823143e-01)(25.000000,8.984740e-02)(30.000000,3.700040e-02)(35.000000,1.386561e-02)(40.000000,5.537633e-03)(45.000000,2.849935e-03)(50.000000,2.005045e-03)(55.000000,1.742446e-03)(60.000000,1.659811e-03)(65.000000,1.633560e-03)(70.000000,1.625124e-03)(75.000000,1.622467e-03)(80.000000,1.621693e-03)
};
   \addplot[smooth, mark=star,only marks,red] plot coordinates {
(0.000000,3.776739e-01)(5.000000,2.978043e-01)(10.000000,1.964127e-01)(15.000000,1.026590e-01)(20.000000,4.274518e-02)(25.000000,1.494992e-02)(30.000000,4.752204e-03)(35.000000,1.500810e-03)(40.000000,5.424761e-04)(45.000000,2.643387e-04)(50.000000,1.813753e-04)(55.000000,1.554188e-04)(60.000000,1.475505e-04)(65.000000,1.450872e-04)(70.000000,1.440788e-04)(75.000000,1.438388e-04)(80.000000,1.437625e-04)
};

\legend{64-QAM, 16-PSK,16-QAM,BPSK,,,,,High SNR,,,,,,,,Simulation};
 \draw \boundellipse{ axis cs:67.5,0.8e-001}{10}{1};
\node at (axis cs:67.5,2.6e-001){\scriptsize{$\xi=0.5$}};

\draw \boundellipse{ axis cs:67.5,1e-003}{10}{2.2};
\node at (axis cs:67.5,1.1e-002){\scriptsize{$\xi=1.1$}};

 \end{axis}
  \end{tikzpicture}
     \caption{Average BER for different modulation schemes under TWTA with ${\rm IBO}=25$ dB and for varying effects of the pointing error with
     $C_n^2(0)=1 \times 10^{-12} {\rm m}^{-\frac{2}{3}}$ with beam wander effect under light shadowing conditions.}
          \label{fig:BER3}
     \end{center}
  \end{figure}
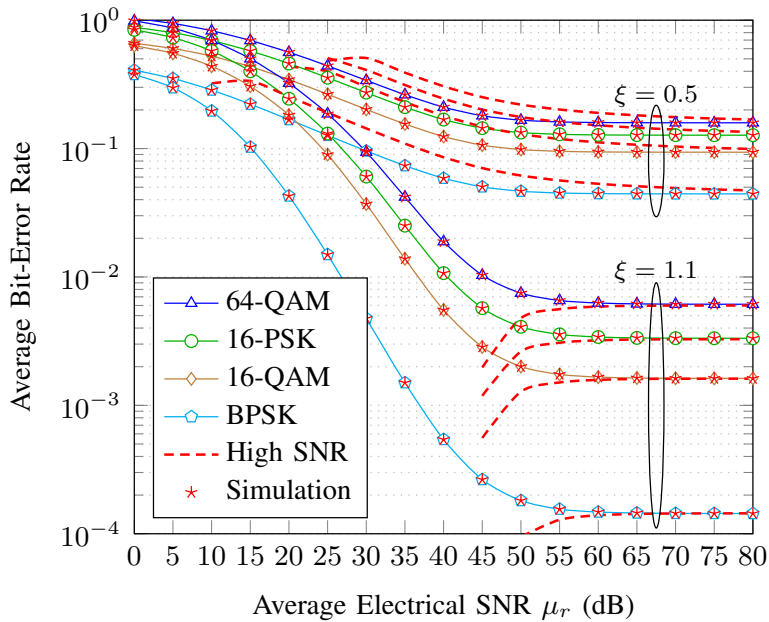

The BER performance for 64-QAM, 16-PSK, 16 QAM, and BPSK modulation schemes under the heterodyne detection technique and TWTA with an IBO of 25 dB, is shown in Fig.~\ref{fig:BER3} with varying effects of the pointing error ($\xi=0.5$ and $1.1$). Clearly, we can observe that the BER performance for all modulation techniques gets better for lower effect of the pointing error (i.e. higher values of $\xi$). Moreover, it can be seen from Fig.~\ref{fig:BER3} that 16-QAM outperforms 16-PSK, as expected when $M > 4$ \cite{proakis2008digital} and BPSK modulation offers the best performance compared to the presented modulation techniques. Other outcomes, particulary for the asymptotic
result at high SNR, can be noticed similar to Fig.~\ref{fig:BER2}.

Fig.~\ref{fig:CAPCITY1} presents the ergodic capacity using the IM/DD technique for negligible effect of the pointing error $\xi=6.7$ in the presence of HPA nonlinearity. Both TWTA and SSPA models are considered with different values of IBO. As clearly seen from the figure, a capacity ceiling is created under both HPA models specially for small values of IBO. Moreover, it can be observed that there is an enhancement in the ergodic capacity as IBO increases and this improvement is greater when SSPA is used. For instance, both SSPA with IBO$=$25 dB and linear PA have the same impact on the ergodic capacity up to 40 dB. This confirms that SSPA performs better than TWTA.
 \begin{figure}[!h]
  \begin{center}
\begin{tikzpicture}[scale=1.2]
    \begin{axis}[font=\footnotesize,
   xlabel= Average Electrical SNR $\mu_r$ (dB), ylabel=  Ergodic Capacity (bps/Hz),
  xmin=0,xmax=80,ymax=15,ymin=0,
    legend style={nodes=right},legend pos= north west,legend style={nodes={scale=1, transform shape}},
      xmajorgrids,
    grid style={dotted},
    ymajorgrids,
   ]

\addplot[smooth,blue,mark=triangle*,mark options={solid},every mark/.append style={solid, fill=white}] plot coordinates {
(0.000000,4.783996e-02)(5.000000,1.342692e-01)(10.000000,3.179466e-01)(15.000000,5.737414e-01)(20.000000,7.812426e-01)(25.000000,8.863389e-01)(30.000000,9.266180e-01)(35.000000,9.403186e-01)(40.000000,9.447806e-01)(45.000000,9.462175e-01)(50.000000,9.466726e-01)(55.000000,9.468197e-01)(60.000000,9.468648e-01)(65.000000,9.468804e-01)(70.000000,9.468861e-01)(75.000000,9.468872e-01)(80.000000,9.468883e-01)
};

 \addplot[smooth,green!70!black,mark=*,mark options={solid},every mark/.append style={solid, fill=white}] plot coordinates {
(0.000000,4.993855e-02)(5.000000,1.460632e-01)(10.000000,3.814876e-01)(15.000000,8.176034e-01)(20.000000,1.353057e+00)(25.000000,1.758065e+00)(30.000000,1.958308e+00)(35.000000,2.034610e+00)(40.000000,2.060489e+00)(45.000000,2.068913e+00)(50.000000,2.071610e+00)(55.000000,2.072471e+00)(60.000000,2.072746e+00)(65.000000,2.072836e+00)(70.000000,2.072862e+00)(75.000000,2.072872e+00)(80.000000,2.072873e+00)
};

\addplot[smooth,blue,mark=triangle*,mark options={solid},every mark/.append style={solid, fill=white}] plot coordinates {
(0.000000,5.094393e-02)(5.000000,1.512139e-01)(10.000000,4.107633e-01)(15.000000,9.623812e-01)(20.000000,1.872856e+00)(25.000000,3.025621e+00)(30.000000,4.138986e+00)(35.000000,4.926718e+00)(40.000000,5.325509e+00)(45.000000,5.481972e+00)(50.000000,5.535856e+00)(55.000000,5.553446e+00)(60.000000,5.559105e+00)(65.000000,5.560908e+00)(70.000000,5.561480e+00)(75.000000,5.561667e+00)(80.000000,5.561721e+00)
};

 \addplot[smooth,green!70!black,mark=*,mark options={solid},every mark/.append style={solid, fill=white}] plot coordinates {
(0.000000,5.103499e-02)(5.000000,1.515962e-01)(10.000000,4.125996e-01)(15.000000,9.717640e-01)(20.000000,1.913804e+00)(25.000000,3.171602e+00)(30.000000,4.555600e+00)(35.000000,5.823665e+00)(40.000000,6.735424e+00)(45.000000,7.217544e+00)(50.000000,7.413093e+00)(55.000000,7.481394e+00)(60.000000,7.503843e+00)(65.000000,7.511070e+00)(70.000000,7.513374e+00)(75.000000,7.514109e+00)(80.000000,7.514341e+00)
};

    \addplot[smooth,blue,mark=triangle*,mark options={solid},every mark/.append style={solid, fill=white}] plot coordinates {
(0.000000,5.106038e-02)(5.000000,1.517029e-01)(10.000000,4.130412e-01)(15.000000,9.737708e-01)(20.000000,1.922300e+00)(25.000000,3.203042e+00)(30.000000,4.656753e+00)(35.000000,6.100417e+00)(40.000000,7.325229e+00)(45.000000,8.137776e+00)(50.000000,8.535781e+00)(55.000000,8.689725e+00)(60.000000,8.742364e+00)(65.000000,8.759513e+00)(70.000000,8.765015e+00)(75.000000,8.766769e+00)(80.000000,8.767329e+00)
};

    \addplot[smooth,green!70!black,mark=*,mark options={solid},every mark/.append style={solid, fill=white}] plot coordinates {
(0.000000,5.108371e-02)(5.000000,1.517647e-01)(10.000000,4.133517e-01)(15.000000,9.749945e-01)(20.000000,1.927152e+00)(25.000000,3.220538e+00)(30.000000,4.714897e+00)(35.000000,6.276201e+00)(40.000000,7.786979e+00)(45.000000,9.091143e+00)(50.000000,1.000069e+01)(55.000000,1.047261e+01)(60.000000,1.066199e+01)(65.000000,1.072784e+01)(70.000000,1.074945e+01)(75.000000,1.075637e+01)(80.000000,1.075861e+01)
};

    \addplot[smooth,cyan,mark=diamond*,mark options={solid},every mark/.append style={solid, fill=white}] plot coordinates {
(0.000000,5.110786e-02)(5.000000,1.518353e-01)(10.000000,4.135501e-01)(15.000000,9.755970e-01)(20.000000,1.929017e+00)(25.000000,3.226861e+00)(30.000000,4.735299e+00)(35.000000,6.340729e+00)(40.000000,7.982922e+00)(45.000000,9.637876e+00)(50.000000,1.129701e+01)(55.000000,1.295719e+01)(60.000000,1.461797e+01)(65.000000,1.627878e+01)(70.000000,1.793981e+01)(75.000000,1.960078e+01)(80.000000,2.126175e+01)
};

\addplot[smooth, mark=star,only marks,red] plot coordinates {
(0.000000,4.783996e-02)(5.000000,1.342692e-01)(10.000000,3.179466e-01)(15.000000,5.737414e-01)(20.000000,7.812426e-01)(25.000000,8.863389e-01)(30.000000,9.266180e-01)(35.000000,9.403186e-01)(40.000000,9.447806e-01)(45.000000,9.462175e-01)(50.000000,9.466726e-01)(55.000000,9.468197e-01)(60.000000,9.468648e-01)(65.000000,9.468804e-01)(70.000000,9.468861e-01)(75.000000,9.468872e-01)(80.000000,9.468883e-01)
};

\addplot[smooth, mark=star,only marks,red] plot coordinates {
(0.000000,4.993855e-02)(5.000000,1.460632e-01)(10.000000,3.814876e-01)(15.000000,8.176034e-01)(20.000000,1.353057e+00)(25.000000,1.758065e+00)(30.000000,1.958308e+00)(35.000000,2.034610e+00)(40.000000,2.060489e+00)(45.000000,2.068913e+00)(50.000000,2.071610e+00)(55.000000,2.072471e+00)(60.000000,2.072746e+00)(65.000000,2.072836e+00)(70.000000,2.072862e+00)(75.000000,2.072872e+00)(80.000000,2.072873e+00)
};
\addplot[smooth, mark=star,only marks,red] plot coordinates {
(0.000000,5.094393e-02)(5.000000,1.512139e-01)(10.000000,4.107633e-01)(15.000000,9.623812e-01)(20.000000,1.872856e+00)(25.000000,3.025621e+00)(30.000000,4.138986e+00)(35.000000,4.926718e+00)(40.000000,5.325509e+00)(45.000000,5.481972e+00)(50.000000,5.535856e+00)(55.000000,5.553446e+00)(60.000000,5.559105e+00)(65.000000,5.560908e+00)(70.000000,5.561480e+00)(75.000000,5.561667e+00)(80.000000,5.561721e+00)
};

\addplot[smooth, mark=star,only marks,red] plot coordinates {
(0.000000,5.103499e-02)(5.000000,1.515962e-01)(10.000000,4.125996e-01)(15.000000,9.717640e-01)(20.000000,1.913804e+00)(25.000000,3.171602e+00)(30.000000,4.555600e+00)(35.000000,5.823665e+00)(40.000000,6.735424e+00)(45.000000,7.217544e+00)(50.000000,7.413093e+00)(55.000000,7.481394e+00)(60.000000,7.503843e+00)(65.000000,7.511070e+00)(70.000000,7.513374e+00)(75.000000,7.514109e+00)(80.000000,7.514341e+00)
};

\addplot[smooth, mark=star,only marks,red] plot coordinates {
(0.000000,5.106038e-02)(5.000000,1.517029e-01)(10.000000,4.130412e-01)(15.000000,9.737708e-01)(20.000000,1.922300e+00)(25.000000,3.203042e+00)(30.000000,4.656753e+00)(35.000000,6.100417e+00)(40.000000,7.325229e+00)(45.000000,8.137776e+00)(50.000000,8.535781e+00)(55.000000,8.689725e+00)(60.000000,8.742364e+00)(65.000000,8.759513e+00)(70.000000,8.765015e+00)(75.000000,8.766769e+00)(80.000000,8.767329e+00)
};

\addplot[smooth, mark=star,only marks,red] plot coordinates {
(0.000000,5.108371e-02)(5.000000,1.517647e-01)(10.000000,4.133517e-01)(15.000000,9.749945e-01)(20.000000,1.927152e+00)(25.000000,3.220538e+00)(30.000000,4.714897e+00)(35.000000,6.276201e+00)(40.000000,7.786979e+00)(45.000000,9.091143e+00)(50.000000,1.000069e+01)(55.000000,1.047261e+01)(60.000000,1.066199e+01)(65.000000,1.072784e+01)(70.000000,1.074945e+01)(75.000000,1.075637e+01)(80.000000,1.075861e+01)
};
\addplot[smooth, mark=star,only marks,red] plot coordinates {
(0.000000,5.110786e-02)(5.000000,1.518353e-01)(10.000000,4.135501e-01)(15.000000,9.755970e-01)(20.000000,1.929017e+00)(25.000000,3.226861e+00)(30.000000,4.735299e+00)(35.000000,6.340729e+00)(40.000000,7.982922e+00)(45.000000,9.637876e+00)(50.000000,1.129701e+01)(55.000000,1.295719e+01)(60.000000,1.461797e+01)(65.000000,1.627878e+01)(70.000000,1.793981e+01)(75.000000,1.960078e+01)(80.000000,2.126175e+01)
};

\legend{TWTA,SSPA,,,,,Linear PA,Simulation};

\draw \boundellipse{ axis cs:67.5,9.8}{10}{13};
\node at (axis cs:67.5,11.5){\scriptsize{IBO$=$25dB}};

\draw \boundellipse{ axis cs:67.5,6.5}{10}{13};
\node at (axis cs:67.5,4.8){\scriptsize{IBO$=$20dB}};

 \draw \boundellipse{ axis cs:67.5,1.5}{10}{9};
\node at (axis cs:67.5,2.8){\scriptsize{IBO$=$10dB}};

 \end{axis}
  \end{tikzpicture}
     \caption{Ergodic capacity under TWTA and SSPA with different values of IBO for $\xi=6.7$ and $C_n^2(0)=1\times 10^{-13} {\rm m}^{-\frac{2}{3}}$ with beam wander effect under light shadowing conditions.}
          \label{fig:CAPCITY1}
     \end{center}
  \end{figure}
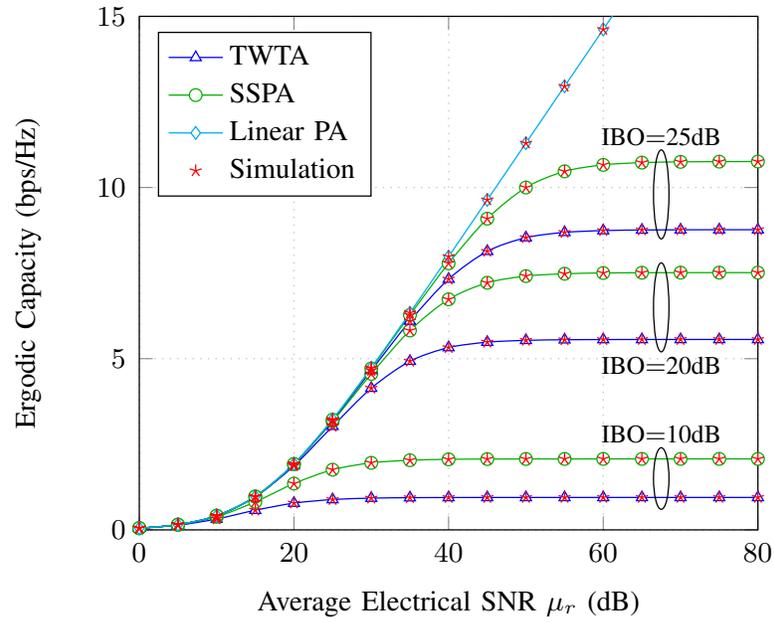\noindent
 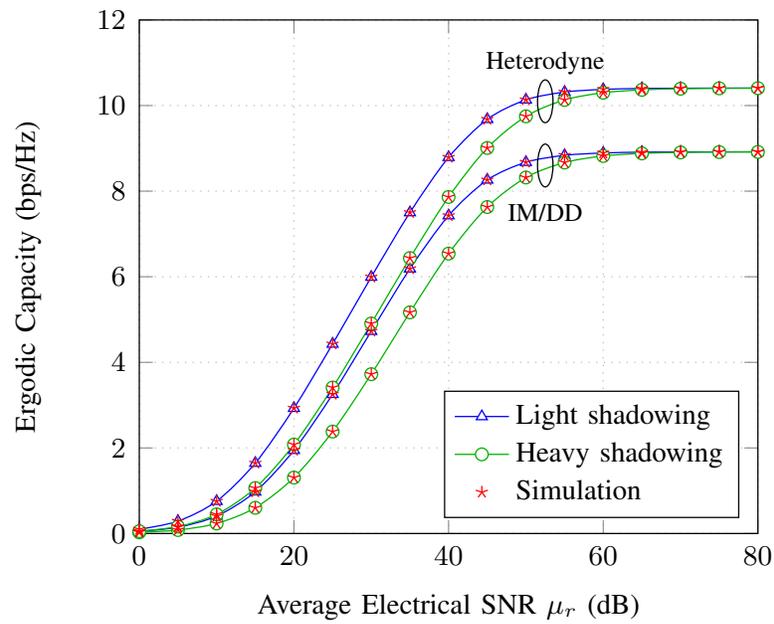
\begin{figure}[!h]
  \begin{center}
\begin{tikzpicture}[scale=1.2]
    \begin{axis}[font=\footnotesize,
   xlabel= Average Electrical SNR $\mu_r$ (dB), ylabel=  Ergodic Capacity (bps/Hz),
  xmin=0,xmax=80,ymax=12,ymin=0,
    legend style={nodes=right},legend pos= south east,legend style={nodes={scale=1, transform shape}},
      xmajorgrids,
    grid style={dotted},
    ymajorgrids,
   ]

\addplot[smooth,blue,mark=triangle*,mark options={solid},every mark/.append style={solid, fill=white}] plot coordinates {
(0.000000,4.888239e-02)(5.000000,1.465942e-01)(10.000000,4.058173e-01)(15.000000,9.742293e-01)(20.000000,1.946007e+00)(25.000000,3.253770e+00)(30.000000,4.727795e+00)(35.000000,6.186303e+00)(40.000000,7.428763e+00)(45.000000,8.262413e+00)(50.000000,8.675616e+00)(55.000000,8.836765e+00)(60.000000,8.891988e+00)(65.000000,8.910021e+00)(70.000000,8.915797e+00)(75.000000,8.917636e+00)(80.000000,8.918224e+00)
};

\addplot[smooth,green!70!black,mark=*,mark options={solid},every mark/.append style={solid, fill=white}] plot coordinates {
(0.000000,2.667197e-02)(5.000000,8.135014e-02)(10.000000,2.341319e-01)(15.000000,6.006381e-01)(20.000000,1.307300e+00)(25.000000,2.381987e+00)(30.000000,3.723149e+00)(35.000000,5.168072e+00)(40.000000,6.540775e+00)(45.000000,7.629045e+00)(50.000000,8.319417e+00)(55.000000,8.668832e+00)(60.000000,8.821552e+00)(65.000000,8.882293e+00)(70.000000,8.905002e+00)(75.000000,8.913447e+00)(80.000000,8.916416e+00)
};

\addplot[smooth,blue,mark=triangle*,mark options={solid},every mark/.append style={solid, fill=white}] plot coordinates {
(0.000000,9.816521e-02)(5.000000,2.878421e-01)(10.000000,7.543908e-01)(15.000000,1.645004e+00)(20.000000,2.927807e+00)(25.000000,4.428950e+00)(30.000000,5.993462e+00)(35.000000,7.500384e+00)(40.000000,8.790779e+00)(45.000000,9.679533e+00)(50.000000,1.013432e+01)(55.000000,1.031532e+01)(60.000000,1.037801e+01)(65.000000,1.039849e+01)(70.000000,1.040508e+01)(75.000000,1.040717e+01)(80.000000,1.040785e+01)
};

\addplot[smooth,green!70!black,mark=*,mark options={solid},every mark/.append style={solid, fill=white}] plot coordinates {
(0.000000,5.377000e-02)(5.000000,1.614841e-01)(10.000000,4.467841e-01)(15.000000,1.061733e+00)(20.000000,2.079018e+00)(25.000000,3.410805e+00)(30.000000,4.907414e+00)(35.000000,6.436413e+00)(40.000000,7.863439e+00)(45.000000,9.009297e+00)(50.000000,9.748569e+00)(55.000000,1.013174e+01)(60.000000,1.030030e+01)(65.000000,1.036747e+01)(70.000000,1.039314e+01)(75.000000,1.040259e+01)(80.000000,1.040614e+01)
};

\addplot[smooth, mark=star,only marks,red] plot coordinates {
(0.000000,4.888239e-02)(5.000000,1.465942e-01)(10.000000,4.058173e-01)(15.000000,9.742293e-01)(20.000000,1.946007e+00)(25.000000,3.253770e+00)(30.000000,4.727795e+00)(35.000000,6.186303e+00)(40.000000,7.428763e+00)(45.000000,8.262413e+00)(50.000000,8.675616e+00)(55.000000,8.836765e+00)(60.000000,8.891988e+00)(65.000000,8.910021e+00)(70.000000,8.915797e+00)(75.000000,8.917636e+00)(80.000000,8.918224e+00)
};

\addplot[smooth, mark=star,only marks,red] plot coordinates {
(0.000000,2.667197e-02)(5.000000,8.135014e-02)(10.000000,2.341319e-01)(15.000000,6.006381e-01)(20.000000,1.307300e+00)(25.000000,2.381987e+00)(30.000000,3.723149e+00)(35.000000,5.168072e+00)(40.000000,6.540775e+00)(45.000000,7.629045e+00)(50.000000,8.319417e+00)(55.000000,8.668832e+00)(60.000000,8.821552e+00)(65.000000,8.882293e+00)(70.000000,8.905002e+00)(75.000000,8.913447e+00)(80.000000,8.916416e+00)
};

\addplot[smooth, mark=star,only marks,red] plot coordinates {
(0.000000,9.816521e-02)(5.000000,2.878421e-01)(10.000000,7.543908e-01)(15.000000,1.645004e+00)(20.000000,2.927807e+00)(25.000000,4.428950e+00)(30.000000,5.993462e+00)(35.000000,7.500384e+00)(40.000000,8.790779e+00)(45.000000,9.679533e+00)(50.000000,1.013432e+01)(55.000000,1.031532e+01)(60.000000,1.037801e+01)(65.000000,1.039849e+01)(70.000000,1.040508e+01)(75.000000,1.040717e+01)(80.000000,1.040785e+01)
};

\addplot[smooth, mark=star,only marks,red] plot coordinates {
(0.000000,5.377000e-02)(5.000000,1.614841e-01)(10.000000,4.467841e-01)(15.000000,1.061733e+00)(20.000000,2.079018e+00)(25.000000,3.410805e+00)(30.000000,4.907414e+00)(35.000000,6.436413e+00)(40.000000,7.863439e+00)(45.000000,9.009297e+00)(50.000000,9.748569e+00)(55.000000,1.013174e+01)(60.000000,1.030030e+01)(65.000000,1.036747e+01)(70.000000,1.039314e+01)(75.000000,1.040259e+01)(80.000000,1.040614e+01)
};

\legend{Light shadowing, Heavy shadowing,,,Simulation};

\draw \boundellipse{ axis cs:52.5,10.1}{10}{5};
\node at (axis cs:52.5,11){\scriptsize{Heterodyne}};

\draw \boundellipse{ axis cs:52.5,8.6}{10}{5};
\node at (axis cs:52.5,7.5){\scriptsize{IM/DD}};

\end{axis}
  \end{tikzpicture}
     \caption{Ergodic capacity under TWTA with ${\rm IBO}=25$ dB for different shadowing conditions
   with $C_n^2(0)=1 \times 10^{-13}$ without beam wander effect under both IM/DD and heterodyne techniques.}
          \label{fig:CAPCITY2}
     \end{center}
  \end{figure}

   In Fig.~\ref{fig:CAPCITY2}, the ergodic capacity performance is plotted for both IM/DD and heterodyne techniques under light as well as heavy shadowing conditions.
 TWTA is considered with 25 dB IBO. It can be observed that as the shadowing conditions get severe, the performance under both types of detection is reduced.
Moreover, the performance of each scheme saturates at the same level for high SNR regardless of the shadowing conditions because of the dominance of the HPA nonlinearity effect. In addition, as shown earlier in the outage performance analysis, the heterodyne technique performs much better than IM/DD under all shadowing conditions.

\section{Conclusion}
In this paper, the performance of a multibeam VHTS system that uses the FSO technology in the feeder link and accounts for HPA
nonlinearity has been analyzed in terms of the outage probability, the average BER, and the ergodic capacity when the FSO link operates under either IM/DD or heterodyne techniques. Closed-form expressions for these performance metrics are obtained in terms of the bivariate Meijer's G function
considering the Gamma-Gamma distribution with beam wander and pointing error effects in the FSO feeder link, and the shadowed Rician fading channel in the RF user link.
In addition, asymptotic results for the outage probability and the average BER in the high SNR regime are derived in terms of simple functions.
The presented numerical results have demonstrated the notable effects of the atmospheric turbulence, the beam wander, the pointing errors, the shadowing conditions, and the HPA nonlinearity on the overall system performance. It has been shown that increasing the transmitted beam size or the nominal ground turbulence levels can result in severe performance degradation because of the increase in the scintillation index. Moreover, pointing errors can significantly degrade the performance, particularly for small values of the pointing error coefficient. Furthermore, the use of the heterodyne detection can considerably reduce the outage probability and the BER and increase the capacity, thereby improving the system performance. Our results also manifested the deleterious effects of the nonlinear distortion introduced by both TWTA and SSPA models compared to the linear PA case, especially with low IBO values, and revealed that the TWTA model leads to the greatest performance degradation.
\appendices
\renewcommand{\theequation}{\thesection.\arabic{equation}}
\setcounter{equation}{0}
\section{CDF of the End-to-End SNDR}\label{appendix:A}
This appendix derives closed-form expression for the CDF of the end-to-end SNDR at the $i$-th UT $\gamma_i$. We start by deriving the CDF of $\Lambda_i= \frac{\gamma_1 \gamma_{2,i}}{\kappa \gamma_{2,i}+C}$  which can be written as
\begin{align}\label{CDFP1}
\nonumber F_{\Lambda_i}(x)&={\rm{Pr}}\left [ \frac{\gamma_1\,\gamma_{2,i}}{\kappa \gamma_{2,i}+C}  \leq   x \right]\\
\nonumber &=1- \int_{0}^{\infty}\left ( 1- {\rm{Pr}}\left [ \frac{\gamma_1\,\gamma_{2,i}}{\kappa \gamma_{2,i}+C}  \leq   x \,|\,\gamma_1 \right]\right )
f_{\gamma_1}(\gamma_1)\,d\gamma_1\\
&=1-\int_{0}^{\infty}\overline{F}_{\gamma_{2,i}}\left ( \frac{C x}{z} \right )f_{\gamma_1}(\kappa \,x +z)\,dz,
\end{align}
where $C=\tr\left [ \left(\bB \bB^{\mbox{\tiny H}} \right)^{-1} \right ]\overline{\gamma}_1+\kappa$ and $\overline{F}_{\gamma_{2,i}}(\cdot)$ stands for the complementary CDF of $\gamma_{2,i}$ derived from (\ref{SNR2integer}) by applying \cite[Eq.(3.351/2)]{Tableofintegrals} as
\begin{align}\label{CCDF2}
\nonumber  \overline{F}_{\gamma_{2,i}}(x)&= \left (\frac{2b_i m_i}{2 b_i m_i+\Omega_i}  \right )^{m_i-1}\exp\left ( -\frac{m_i \,x}{\overline{\gamma}_{2,i}} \right )\\
&\times \sum_{k=0}^{m_i-1}\frac{(-1)^k (1-m_i)_k}{k!} \left (\frac{\Omega_i}{2 b_i m_i}  \right )^k
\sum_{j=0}^{k}\frac{1}{j!}\left (  \frac{m_i\, x}{\overline{\gamma}_{2,i}} \right )^j.
\end{align}
Substituting (\ref{CCDF2}) and (\ref{SNR1}) into (\ref{CDFP1}), transforming the $\exp(\cdot)$ function to its correspondent Meijer's G function by applying \cite[Eq.(01.03.26.0004.01)]{Wolfram}, using the definition of the Meijer's G function given in \cite[Eq.(9.301)]{Tableofintegrals}, and interchanging the integrals, the CDF of $\Lambda$ becomes
\begin{align}\label{CDFP2}
\nonumber  F_{\Lambda_i}(x)&= 1-\frac{\xi^2}{r\, \Gamma(\alpha)\Gamma(\beta)}\left (\frac{2b_i m_i}{2 b_i m_i+\Omega_i}  \right )^{m_i-1}\\
\nonumber & \times \sum_{k=0}^{m_i-1}\sum_{j=0}^{k}\frac{(-1)^k (1-m_i)_k}{k!\,j!} \left (\frac{\Omega_i}{2 b_i m_i}  \right )^k\left (  \frac{Cm_i\, x}{\overline{\gamma}_{2,i}} \right )^j \frac{1}{(2\pi i)^{2}}\\
\nonumber & \times \int\limits_{\mathcal{L}_1}\int\limits_{\mathcal{L}_2}\Gamma(-s)\left ( \frac{C m_i x}{\overline{\gamma}_{2,i}} \right )^s \frac{\Gamma(\xi^2-t)\Gamma(\alpha-t)\Gamma(\beta-t)}{\Gamma(\xi^2+1-t)}\\
& \times \left (\frac{\alpha \beta \xi^2}{(\xi^2+1)\mu_r^{\frac{1}{r}}}  \right )^t
\int_{0}^{\infty}z^{-j-s}(z+\kappa\, x)^{\frac{t}{r}-1}\,dz \,ds\, dt,
\end{align}
where $\mathcal{L}_1$ and $\mathcal{L}_2$ represent the $s$- and $t$-plane contours, respectively. Utilizing \cite[Eq.(3.194/3)]{Tableofintegrals} then \cite[Eq.(8.384/1)]{Tableofintegrals}, $\int_{0}^{\infty}z^{-j-s}(z+\kappa\, x)^{\frac{t}{r}-1}\,dz $ reduces to $(\kappa\,x)^{\frac{t}{r}-s-j}\Gamma(1-j-s)\Gamma(j+s-\frac{t}{r})/\Gamma(1-\frac{t}{r})$, and (\ref{CDFP2}) can be re-written as
\begin{align}\label{CDFP3}
\nonumber F_{\Lambda_i}(x)&= 1-\frac{\xi^2}{\Gamma(\alpha)\Gamma(\beta)}\left (\frac{2b_i m_i}{2 b_i m_i+\Omega_i}  \right )^{m_i-1} \\
\nonumber & \times \sum_{k=0}^{m_i-1}\sum_{j=0}^{k}\frac{(-1)^k (1-m_i)_k}{k!\,j!}\left (\frac{\Omega_i}{2 b_i m_i}  \right )^k
\frac{1}{(2\pi i)^{2}}
\int\limits_{\mathcal{L}_1}\int\limits_{\mathcal{L}_2}\Gamma(s+t)\\
\nonumber & \times \Gamma(j-s) \Gamma(1-s)\left ( \frac{C m_i}{\kappa\,\overline{\gamma}_{2,i}} \right )^s \frac{\Gamma(\xi^2+rt)\Gamma(\alpha+rt)\Gamma(\beta+rt)}{\Gamma(\xi^2+1+rt)\Gamma(1+t)}\\
&\times \left (\frac{(\xi^2+1)^r \mu_r }{(\alpha \beta \xi^2)^r \kappa \,x}  \right )^t\,ds \,dt.
\end{align}
Plugging $\Gamma(n z)=n^{n z-\frac{1}{2}} (2 \pi)^{\frac{1-n}{2}}\prod_{k=0}^{n-1}\Gamma\left ( z+\frac{k}{n} \right )$ for $n \in \mathbb{N}$ in (\ref{CDFP3}) yields
\begin{align}\label{CDFP4}
\nonumber  F_{\Lambda_i}(x)&= 1-\frac{\xi^2\, r^{\alpha+\beta-2}}{\Gamma(\alpha)\Gamma(\beta) (2 \pi)^{r-1}}\left (\frac{2b_i m_i}{2 b_i m_i+\Omega_i}  \right )^{m_i-1}\sum_{k=0}^{m_i-1}\sum_{j=0}^{k} \\
\nonumber &  \times \frac{(-1)^k (1-m_i)_k}{k!\,j!}\left (\frac{\Omega_i}{2 b_i m_i}  \right )^k
\frac{1}{(2\pi i)^{2}}\int\limits_{\mathcal{L}_1}\int\limits_{\mathcal{L}_2}\Gamma(s+t)\Gamma(j-s) \\
\nonumber & \times \Gamma(1-s)\left ( \frac{C m_i}{\kappa\,\overline{\gamma}_{2,i}} \right )^s \frac{\prod_{i=0}^{r-1}\Gamma\left(\frac{\xi^2+i}{r}+t\right)\prod_{i=0}^{r-1}\Gamma\left(\frac{\alpha+i}{r}+t\right)}{\prod_{i=0}^{r-1}\Gamma\left(\frac{\xi^2+1+i}{r}+t\right)\Gamma(1+t)}\\
& \times \prod_{i=0}^{r-1}\Gamma\left(\frac{\beta+i}{r}+t\right) \left (\frac{r^{2r}(\xi^2+1)^r \mu_r }{(\alpha \beta \xi^2)^r \kappa \,x}  \right )^t\,ds \,dt.
\end{align}
With the help of \cite[Eq.(1)]{Gupta}, the CDF of $\Lambda_i$ can be derived in terms of the bivariate Meijer's G function as
\begin{align}\label{CDFP5}
\nonumber  F_{\Lambda_i}(x)&= 1-\frac{\xi^2\, r^{\alpha+\beta-2}}{\Gamma(\alpha)\Gamma(\beta) (2 \pi)^{r-1}}\left (\frac{2b_i m_i}{2 b_i m_i+\Omega_i}  \right )^{m_i-1} \\
\nonumber & \times \sum_{k=0}^{m_i-1}\sum_{j=0}^{k}\frac{(-1)^k (1-m_i)_k}{k!\,j!}\left (\frac{\Omega_i}{2 b_i m_i}  \right )^k \\
&\times {\rm{G}}_{1,0:0,2:3r,r+1}^{1,0:2,0:0,3r}\begin{bmatrix}
\begin{matrix}
0
\end{matrix}
\Bigg|\begin{matrix}
-\\j,1
\end{matrix}
\Bigg|\begin{matrix}
\mathcal{K}_1\\\Delta (r,-\xi^2),0
\end{matrix}
\Bigg|
\frac{C m_i}{\kappa\,\overline{\gamma}_{2,i}},\frac{r^{2r}(\xi^2+1)^r \mu_r }{(\alpha \beta \xi^2)^r \kappa \,x}
\end{bmatrix},
\end{align}
where $\mathcal{K}_1=\Delta (r,1-\xi^2),\Delta (r,1-\alpha),\Delta (r,1-\beta)$ and $\Delta (r,u)=\frac{u}{r},\frac{u+1}{r},\ldots,\frac{u+r-1}{r}$.
Finally, the desired CDF expression of the end-to-end SNDR at the $i$-th UT, $\gamma_i$, can be easily obtained from (\ref{CDFP5}) using a simple RV transformation as shown by (\ref{CDFtotal}).
\setcounter{equation}{0}
\section{Moments}\label{appendix:B}
By applying \cite[Eq.(12)]{Gupta}, the bivariate Meijer's G function in (\ref{PDFtotal}) can be written as a definite integral involving the product of three Meijer's G functions and therefore, the moments can be expressed as
\begin{align}\label{moments1}
\nonumber  \mathbb{E}[\gamma_i^n]&=\frac{\xi^2\, r^{\alpha+\beta-2}}{\Gamma(\alpha)\Gamma(\beta)(2 \pi)^{r-1}}
\left (\frac{2b_i m_i}{2 b_i m_i+\Omega_i}  \right )^{m_i-1} \sum_{k=0}^{m_i-1}\sum_{j=0}^{k}\\
\nonumber & \times \frac{(-1)^k (1-m_i)_k}{k!\,j!}\left (\frac{\Omega_i}{2 b_i m_i}  \right )^k
\int_{0}^{\infty}\frac{e^{-z}}{z}{\rm{G}}_{0,2}^{2,0}\left[\frac{C\,m_i z}{\kappa\overline{\gamma}_{2,i}}\left| \begin{matrix} {-} \\ {j,1} \\ \end{matrix} \right. \right]\\
& \times \int_{0}^{\infty} x^{n-1}{\rm{G}}_{3r,r+1}^{0,3r}\left[\frac{r^{2r}(\xi^2+1)^r \mu_r z}{(\alpha \beta \xi^2)^r \kappa \left \| \bb_i^{\mbox{\tiny T}} \right \|^2 x} \left| \begin{matrix} {\mathcal{K}_1}\\{\Delta (r,-\xi^2),1} \\ \end{matrix} \right. \right]dx\, dz.
\end{align}
By transforming the Meijer's G function to its correspondent Fox's H function with the help of \cite[Eq. (2.9.1)]{HTranforms}, inverting the argument of the obtained Fox's H function via \cite[Eq. (2.1.3)]{HTranforms}, applying \cite[Eq. (2.4.5)]{HTranforms}, and utilizing the integral identity \cite[Eq. (2.8)]{HTheory}, (\ref{moments1}) reduces to

\begin{align}\label{moments2}
\nonumber  \mathbb{E}[\gamma_i^n]&=\frac{\xi^2\Gamma(\alpha+r\,n)\Gamma(\beta+ r\,n)}{(\xi^2+r\,n)\Gamma(\alpha)\Gamma(\beta)\Gamma(n)}\left (\frac{(\xi^2+1)^r \mu_r}{(\alpha \beta \xi^2)^r \kappa  \left \| \bb_i^{\mbox{\tiny T}} \right \|^2 }  \right )^n\\
&\times\left (\frac{2b_i m_i}{2 b_i m_i+\Omega_i}  \right )^{m_i-1}
\int_{0}^{\infty}z^{n-1}e^{-z}{\rm{G}}_{0,2}^{2,0}\left[\frac{Cm_i z}{\kappa\overline{\gamma}_{2,i}}\left| \begin{matrix} {-} \\ {j,1} \\ \end{matrix} \right. \right]dz.
\end{align}
Finally, (\ref{moments2}) is further simplified to (\ref{moments}) by exploiting the integral identity \cite[Eq.(7.813)]{Tableofintegrals}.
\setcounter{equation}{0}

\bibliographystyle{IEEEtran}
\bibliography{IEEEexample}
\end{document}